\documentclass[english,aps,reprint,twocolumn,
  superscriptaddress,
  longbibliography] {revtex4-1}
\usepackage[english]{babel} 

\usepackage{amsmath,amsfonts,amssymb}
\usepackage{array}
\usepackage{multirow}
\usepackage{graphicx}
\usepackage{color}
\usepackage{paralist}
\usepackage{hyperref}

\usepackage{placeins} 

\usepackage[normalem]{ulem}
\usepackage{array,mathtools}
\usepackage{enumitem}
\usepackage{xcolor,colortbl}
 \definecolor{myred}{HTML}{8B0000}
 \definecolor{mygreen}{HTML}{006400}
 \definecolor{myblue}{HTML}{00008B}
 \definecolor{myyellow}{HTML}{FFD700}
 \definecolor{myindigo}{HTML}{4B0082}
 \definecolor{myorange}{HTML}{D2691E}
 \definecolor{mygray}{HTML}{808080}
\usepackage{booktabs}
 \AtBeginDocument{\heavyrulewidth=.08em
                  \lightrulewidth=.05em
                  \cmidrulewidth=.03em
                  \belowrulesep=.65ex
                  \belowbottomsep=0pt
                  \aboverulesep=.4ex
                  \abovetopsep=0pt
                  \cmidrulesep=\doublerulesep
                  \cmidrulekern=.5em
                  \defaultaddspace=.5em
                  }
 \newcolumntype{L}[1]{>{\raggedright\arraybackslash}p{#1}}
\usepackage{tikz}
 \usetikzlibrary{arrows,shapes,fit,calc,positioning,spy,decorations.pathreplacing,decorations.pathmorphing,patterns,backgrounds}
 \tikzset{gene/.style={draw,circle,minimum size=1.05em},
          protein/.style={draw,regular polygon,regular polygon sides=3,minimum size=4em,yshift=-0.2em,scale=0.3},
          compound/.style={draw,regular polygon,regular polygon sides=4,minimum size=1.35em},
          reaction/.style={draw,regular polygon,regular polygon sides=6,minimum size=1.04em},
          encodes/.style={-latex},
          activates/.style={-latex,dashed,shorten >=0.1em}, 
          inhibites/.style={-|, dashed,shorten >=0.1em},
          box/.style={draw=JUlblue,rounded corners=5mm,inner sep=5mm,ultra thick,dashed,fill=JUlblue!60},
          node/.style={draw,JUdblue,fill,circle,inner sep=0.2em},
          edge/.style={-latex,thick,mygray},
          line/.style={dashed,thick,JUdblue},
          Square/.style={draw, rectangle, thick, minimum height=2.5em, minimum width=6em},
          square/.style={draw, rectangle, minimum height=2em, minimum width=3em},
          rhombus/.style={draw, diamond, aspect=2, inner sep=0mm, minimum size=1.5em},
          Box/.style={draw, thick, inner sep=2mm}
          }

\begin{document}
\title{The interdependent network of gene regulation and metabolism
  is robust where it needs to be}
\author{David F. Klosik}
\email{klosik@itp.uni-bremen.de}
\affiliation{Institute for Theoretical Physics, University of Bremen}
\author{Anne Grimbs}
\email{a.grimbs@jacobs-university.de}
\affiliation{Department of Life Sciences and Chemistry, Jacobs University Bremen}
\author{Stefan Bornholdt}
\email[\textbf{corresponding author:} ]{bornholdt@itp.uni-bremen.de}
\affiliation{Institute for Theoretical Physics, University of Bremen}
\author{Marc-Thorsten H\"utt}
\email{m.huett@jacobs-university.de}
\affiliation{Department of Life Sciences and Chemistry, Jacobs University Bremen}

\begin{abstract} 
The major biochemical networks of the living cell, the network of
interacting genes and the network of biochemical reactions, are
highly interdependent, however, they have been studied mostly as
separate systems so far.  In the last years an appropriate
theoretical framework for studying interdependent networks has been
developed in the context of statistical physics.

Here we study the interdependent network of gene regulation and
metabolism of the model organism \textit{Escherichia coli} using the
theoretical framework of interdependent networks.

In particular we aim at understanding how the biological system 
can consolidate the conflicting tasks of reacting rapidly to 
(internal and external) perturbations, while being robust to minor 
environmental fluctuations, at the same time. For this purpose 
we study the network response to localized perturbations and find  
that the interdependent network is sensitive to gene regulatory 
and protein-level perturbations, yet robust against metabolic changes.

This first quantitative application of the theory of interdependent 
networks to systems biology shows how studying network responses to
localized perturbations can serve as a useful strategy for analyzing 
a wide range of other interdependent networks.
\end{abstract}

\maketitle

A main conceptual approach of current research in the life sciences
is to advance from a detailed analysis of individual molecular
components and processes towards a description of biological
\textit{systems} and to understand the emergence of biological
function from the interdependencies on the molecular level.

Supported by the diverse high-throughput 'omics' technologies, the
relatively recent discipline of systems biology has been the major
driving force behind this new perspective which becomes manifest, for 
example, in the effort to compile extensive databases of biological 
information to be used in genome-scale models 
\cite{kitano_systems_2002,ideker_new_2001,aderem_systems_2005}.

Despite its holistic 'game plan', however, systems biology 
frequently operates on the level of subsystems: Even when 
considering cell-wide transcriptional regulatory networks, 
as, e.g., in a network motif analysis \cite{milo_network_2002}, 
this is only one of the cell's networks. Likewise, the popular 
approach to studying metabolic networks in systems biology, 
constraint-based modeling, accounts for steady-state predictions 
of metabolic fluxes of genome-scale metabolic networks 
\cite{llaneras_stoichiometric_2008}, which again, is only one 
of the other networks of the cell. 

In the analysis of such large networks, systems biology draws 
its tools considerably from the science of complex networks 
which provides a mathematical framework especially suitable 
for addressing interdisciplinary questions. 
Combining the mathematical subdiscipline of graph theory with 
methods from statistical physics, this new field greatly contributed 
to the understanding of, e.g., the percolation properties of networks 
\cite{cohen_percolation_2002}, potential processes of network formation
\cite{dorogovtsev_evolution_2002} or the spreading of disease on
networks \cite{newman_spread_2002}.
In the early 2000s, gene regulation and metabolism have been among
the first applications of the formalisms of 'network biology'
\cite{Barabasi:2004p17908}. Among the diverse studies of network
structure for these systems, the most prominent ones on the gene
regulatory side are the statistical observation and functional
interpretation of small over-represented subgraphs ('network motifs')
\cite{Shen-Orr:Nat-Genet2002,Alon:2007p17924} and the hierarchical
organization of gene regulatory networks \cite{yu2006genomic}. On the
metabolic side, the broad degree distribution of metabolic networks
stands out \cite{Jeong:2000:Nature:11034217}, with the caveat,
however, that 'currency metabolites' (like ATP or H$_2$O) can
severely affect network properties \cite{ma2003reconstruction}, as
well as the hierarchical modular organization of metabolic networks
\cite{Ravasz:2002p351,Guimera:2005p15058}.

Over the last decade, the field of complex networks
moved its focus from the investigation of single-network
representations of systems to the interplay of networks that interact
with and/or depend on each other. Strikingly, it turned out that 
explicit interdependence between network constituents can
fundamentally alter the percolation properties of the resulting
\textit{interdependent networks}, which can show a discontinuous
percolation transition in contrast to the continuous behavior in
single-network percolation 
\cite{parshani_critical_2011,parshani_interdependent_2010,
son_percolation_2012,radicchi_abrupt_2013,
zhou_simultaneous_2014,radicchi_percolation_2015}.
It has also been found that, contrary to the isolated-network case,
networks with broader degree distribution become remarkably fragile
as interdependent networks \cite{buldyrev_catastrophic_2010}.

However, this set of recent developments in network science still 
lacks application to systems biology. 

Arguably, the most prominent representative of interdependent
networks in a biological cell is the combined system of gene
regulation and metabolism which are interconnected by various forms
of protein interactions, e.g.,~enzyme catalysis of biochemical
reactions couples the regulatory to metabolic network, while the
activation or deactivation of transcription factors by certain
metabolic compounds provides a coupling in the opposite direction.

Although it is well-known that gene regulatory and metabolic
processes are highly dependent on one another only few studies
addressed the interplay of gene regulation and metabolism on a larger
scale and from a systemic perspective 
\cite{covert_integrating_2004,shlomi_genome-scale_2007,samal_regulatory_2008}.
The first two studies have aimed at finding consistent metabolic-regulatory 
steady states by translating the influence of metabolic processes
on gene activity into metabolic flux predicates and incorporating
high-throughput gene expression data. This can be considered as an
extension of the constraints-based modeling framework beyond the 
metabolic network subsystem. 
In the paper of Samal and Jain \cite{samal_regulatory_2008}, 
on the other hand, the transcriptional regulatory network of
\textit{Escherichia coli} (\textit{E.~coli}) metabolism has been studied as 
a Boolean network model into which flux predicates can be included as
additional interactions. These models were first important steps 
towards integrating the subsystems of gene regulation and metabolism 
from a systems perspective. 

The formalism of interdependent networks now allows us to go beyond 
these pioneering works on integrative models, by analyzing the  
robustness of the combined system in terms of the maximal effect 
a small perturbation can have on such interdependent systems. 
In particular, the findings can be interpreted in the context 
of cascading failures and percolation theory.

We here undertake a first application of the new methodological 
perspective to the combined networks 
of gene regulation and metabolism in \textit{E.~coli}. 

Using various biological databases, particularly \mbox{EcoCyc} as the
main core \cite{keseler_ecocyc:_2005,keseler_ecocyc:_2013}, we have
compiled a graph representation of gene regulatory and metabolic
processes of \textit{E.~coli} including a high level of detail in the
structural description, distinguishing between a comparatively large
number of node and link types according to their biological
functionality. 

A structural analysis of this compilation reveals that, in addition
to a small set of direct links, the gene-regulatory and the metabolic
domains are predominantly coupled via a third network domain
consisting of proteins and their
interactions. Figure~\ref{fig:sketch} shows this three-domain
functional division. Details about the data compilation, the network
reconstruction and the domain-level analysis are given in Grimbs
\textit{et al.~}\cite{grimbs_integrated_2016}.

This rich structural description, together with purpose-built,
biologically plausible propagation rules allows us to assess the
functional level with methods derived from percolation theory.  More
precisely, we will investigate cascading failures in the three-domain
system, emanating from small perturbations, localized in one of the
domains. By network response to localized perturbations analysis we
will observe below that (i) randomized versions of the graph are much
less robust than the original graph and (ii) that the integrated
system is much more susceptible to small perturbations in the gene
regulatory domain than in the metabolic one.

\section{The System\label{sec:system}}

\begin{figure}[h]
  \centering
  \includegraphics[width=.85\columnwidth]
                  {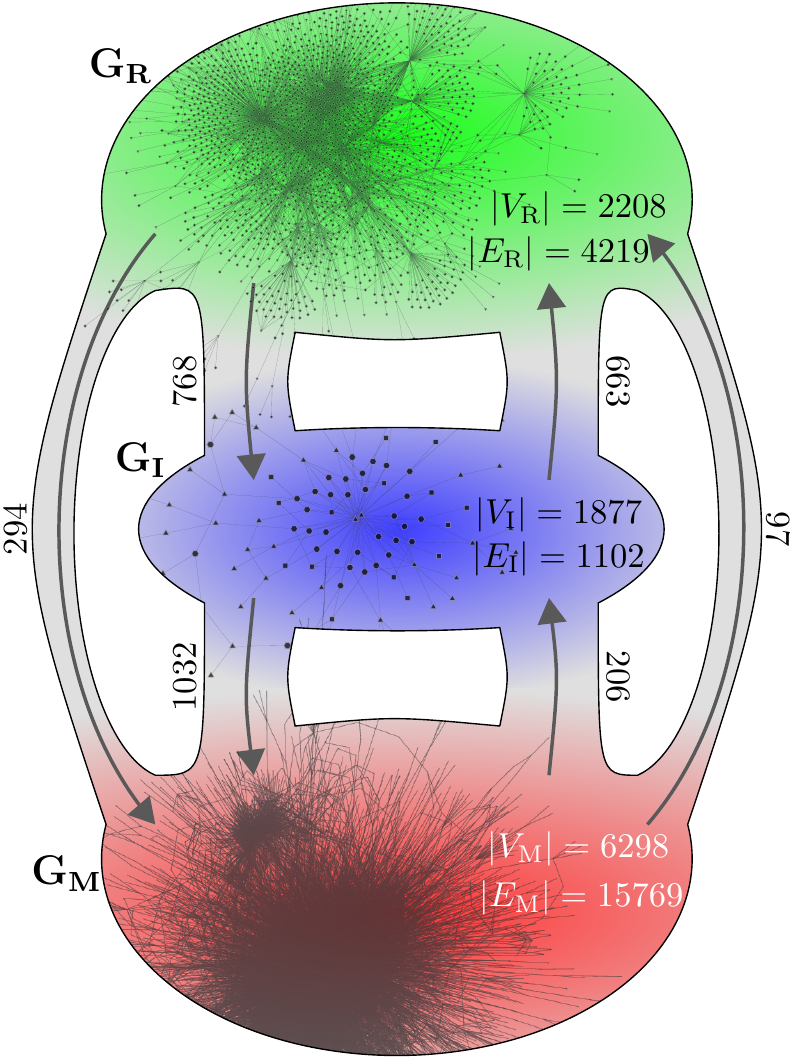}
  \caption{\label{fig:sketch}A sketch of the domain organization of
    the integrative \textit{E.~coli} network. The regulatory domain,
    $G_R$, at the top is connected to the metabolic domain, $G_M$,
    shown at the bottom via a protein-interface layer, $G_I$. The
    figure shows the number of vertices within and the number of
    edges within and between the domains. For illustrative purposes
    snapshots of the largest weakly-connected components of $G_R$,
    $G_I$ and $G_M$ are included in the figure.}
\end{figure}

The core object of our investigation is an \textit{E.~coli} network
representation of its combined gene regulation and metabolism, which
can be thought of as functionally divided into three
\textit{domains}: the representation captures both gene regulatory
and metabolic processes, with these processes being connected by an
intermediate layer that models both, the enzymatic influence of genes
on metabolic processes, as well as signaling-effects of the metabolism
on the activation or inhibition of the expression of certain genes.
The underlying interaction graph $G = (V, E) = \{G_R, G_I, G_M\}$
with its set of nodes (vertices) $V$ and links (edges) $E$ consists
of three interconnected subgraphs, the gene regulatory domain $G_R$,
the interface domain $G_I$ and the metabolic domain $G_M$. From the
functional perspective, $G$ is the union of gene regulatory ($G_R$)
and metabolic processes ($G_M$), and their interactions and
preparatory steps form the interface ($G_I$). Figure~\ref{fig:sketch}
shows a sketch of the network model.
The integrated network representation has been assembled based on the
\mbox{EcoCyc} database (version 18.5;
\cite{keseler_ecocyc:_2005,keseler_ecocyc:_2013}) which offers both
data about metabolic processes and (gene) regulatory events
incorporated from the corresponding RegulonDB release 8.7
\cite{salgado_regulondb_2013}. The extensive metadata allows for the
assignment of the vertices to one of the three functional
domains. Details of this process and a detailed characterization of
the resulting model will be described elsewhere
\cite{grimbs_integrated_2016}. The corresponding graph representation
consists of $10,383$ vertices and $24,150$ directed edges.

Since we are interested in the propagation of a signal between the
domains, in the following we will refer to the domains of the source
and target vertices of the edge $e_i=(v_s^{(i)},v_t^{(i)})$ as
\textit{source domain}, SD, and \textit{target domain}, TD,
respectively. The metadata can be used to assign properties to the
nodes and edges of the graph beyond the domain structure, some of
which are used in the following analysis, namely in the construction
of the propagation rules of the system and of the randomization
schemes.

We distinguish between biological categories of edges (capturing the 
diverse biological roles of the edges) and the logical categories 
(determining the rules of the percolation process). 
According to their biological role in the system, both vertices and
edges are assigned to a \textit{biological} category; we
abbreviate the \textit{biological category of a vertex} as BCV and
the \textit{biological category of an edge} as BCE (for details see 
Supplementary Materials).
Each of the eight BCEs can then be mapped uniquely to one of only
three \textit{logical categories of an edge}, LCE, 
\begin{displaymath}
e_i^\text{LCE} \in  \{C, D, R \}
\end{displaymath}
which are of central importance for the spreading dynamics in our
system:

\begin{description}
  \item[C, 'conjunct'] The target vertex of an edge with this logical
    \texttt{AND} property depends on the source node, i.e., it will
    fail once the source node fails. For example, for a reaction to
    take place, all of its educts have to be available.
  \item[D, 'disjunct'] Edges with this logical \texttt{OR} property
    are considered redundant in the sense that a vertex only fails
    if the source vertices of all of its incoming D-edges fail. For
    instance, a compound will only become unavailable once all of
    its producing reactions have been canceled.
  \item[R, 'regulation'] Edges of this category cover $14$
    different kinds of regulatory events (described in detail in
    the Supplementary Materials). As shown below, in terms of the
    propagation dynamics we treat these edges similar to the
    'conjunct' ones.
\end{description}

\section{Perturbation Rules\label{sec:updates}}

Next we describe the dynamical rules for the propagation of an
initial perturbation in the network in terms of the logical 
categories of an edge (LCE), which distinguish between the 
different roles a given edge has in the update of a target vertex.

Every vertex is assigned a Boolean state variable
\mbox{$\sigma\in\{0,1\}$}; since we intend to mimic the propagation
of a perturbation (rather than simulate a trajectory of actual
biological states) we identify the state $1$ with \textit{not yet
affected by the perturbation} while the state $0$ corresponds to
\textit{affected by the perturbation}. We stress that the trajectory
$\vec{\sigma}(t)$ does not correspond to the time evolution of the
abundance of gene products and metabolic compounds, but the rules
have been chosen such that the final set of affected nodes provides
an estimate of all nodes potentially being affected by the initial
perturbation. A node not in this set is topologically very unlikely
of being affected by the perturbation at hand (given the biological
processes contained in our model).

A stepwise update can now be defined for vertex $i$ with
in-neighbours $\Gamma^{-}_{i}$ in order to study the spreading of
perturbations through the system by initially switching off a
fraction $q$ of vertices:
\begin{align}
  & \sigma_i(t+1) = f_{i} ( \sigma_j(t) | \sigma_j \in \Gamma^{-}_{i} ) \\
  & f_i = \begin{cases} 1 & 
    \mbox{if }
    \parbox[t]{.5\columnwidth}{$
      \left[\sum_j c_{ij}
      (1-\sigma_j) = 0\right] \wedge \; \left[\sum_j d_{ij}
      \sigma_j > 0 \vee \sum_j d_{ij}=0 \right] \;\wedge \;
      \left[\sum_j |r_{ij}| (1-\sigma_j) = 0\right]
    $}
    \\ 0 &
    \text{otherwise}
  \end{cases}
\end{align}
where $c_{ij}$ is $1$ if $v_j$ is connected to $v_i$ via a C-link and
$0$ otherwise; $d_{ij}$ and $r_{ij}$ are defined analogously.

Thus, a vertex will be considered unaffected by the perturbation if
none of its in-neighbours connected via either a $C$ or an $R$ edge
have failed (regardless of the sign of the regulatory interaction),
and at least one of its in-neighbours connected via a $D$ edge is
still intact. With an additional rule it is ensured that an initially 
switched off vertex stays off.
The choice of the update rules ensures that the unperturbed system
state is conserved under the dynamics: $\vec{f}(\vec{1})=\vec{1}$.

As a side remark, the spreading of a perturbation according to the
rules defined above could also be considered as an epidemic process
with one set of connections with a very large, and a second set of
connections with a very low probability of infection
\cite{watts_simple_2002}.

\section{Elements of Percolation Theory}
In systems which can be described without explicit dependencies
between its constituents but with a notion of functionality that
coincides with connectivity, percolation theory is a method of first
choice to investigate the system's response to average perturbations
of a given size that can be modelled as failing vertices or edges
\cite{callaway_network_2000,cohen_percolation_2002}
The fractional size of the giant connected component as a function of
the occupation probability $p$ of a constituent typically vanishes at
some critical value $p_c$, the percolation threshold. In the
following, we will mostly use the complementary quantity $q=1-p$ so
that $q_c=1-p_c$ can be interpreted as the critical size of the
initial attack or perturbation of the system.
The strong fluctuations of the system's responses in the vicinity of
this point can serve as a proxy for the percolation threshold, which
is especially useful in finite systems in which the transition
appears smoothed out. In our analysis, the susceptibility
$\tilde{\chi} = \langle S^2 \rangle - \langle S \rangle^2$, where $S$
is the size of the largest cluster, is used
\cite{PhysRevE.93.030302}.

Upon the introduction of explicit dependencies between the system's
constituents, the percolation properties can change dramatically. The
order parameter no longer vanishes continuously but typically jumps
at $p_c$ in a discontinuous transition
\cite{parshani_interdependent_2010,parshani_critical_2011} as
cascades of failures fragment the system. A broader degree
distribution now enhances a graph's vulnerability to random failures,
in opposition to the behavior in isolated graphs
\cite{buldyrev_catastrophic_2010,albert_error_2000}. 
Details of the corresponding theoretical framework have been worked
out by Parshani \textit{et al.~}\cite{parshani_interdependent_2010},
Son \textit{et al.~}\cite{son_percolation_2012}, Baxter \textit{et al.~}\cite{baxter_avalanche_2012}
and more recently the notion of 'networks of networks' has been
included
\cite{gao_networks_2011,gao_single_2014,kenett_networks_2015}. There
have also been attempts to integrate this class of models into the
framework of multilayer networks \cite{kivela_multilayer_2014}.

In addition to random node failure other procedures for initial node
removal have been explored, e.g.,~node removal with respect to their
degree (targeted attacks) \cite{huang_robustness_2011}.

Currently, two notions of \textit{localized} attacks have been
described. Attacks of the first sort are defined on spatially
embedded networks and are 'local' with respect to a distance in this
embedding, i.e.~in a 'geographical' sense \cite{berezin_localized_2015}.  
The second approach considers locality in terms of connectivity: 
around a randomly chosen seed, neighbours are removed layer by layer
\cite{yuan_how_2015,shao_percolation_2015}.
In contrast, as described below in our approach, attacks are
localized with respect to the three network domains, while within the
domains nodes are chosen randomly.

At this point we would like to shortly comment on the applicability 
of the mathematical concepts of interdependent networks to
real-world data. Aiming at analytical tractability, typical model
systems need to choose a rather high level of abstraction. While
certainly many systems can be accurately addressed in that way, we
argue that especially in the case of biological systems the
theoretical concepts can require substantial adjustment to cover
essential properties of the system at hand. 
 
When asking for the systemic consequences of interdependency, the
distinction between several classes of nodes and links may be
required. Effectively, some classes of links may then represent
simple connectivity, while others can rather be seen as dependence
links. In Biology, such dependencies are typically mediated by
specific molecules (e.g., a small metabolite affecting a
transcription factor, or a gene encoding an enzyme catalyzing a
biochemical reaction). Such implementations of dependence links are
no longer just associations and it is hard to formally distinguish
them from the functional links.

In contrast to the explicitly alternating 'percolation' and
'dependency' steps in typical computational models in which 
the decoupling of nodes from the largest component yields 
dependent nodes to fail, in our directed model both, 
connectivity and dependency links are evaluated
in every time step and (apart from nodes failing due to dependency)
only fully decoupled vertices cause further dependency failures.

\section{Network Response to Localized Perturbation Analysis\label{sec:nrlp}}

Due to the functional three-domain partition of our \textit{E.~coli} 
gene regulatory and metabolic network
reconstruction, we have the possibility to classify perturbations not
only according to their size, but also with respect to their
localization in one of the domains comprising the full interdependent
system, thereby enabling us to address the balance of sensitivity and
robustness of the interdependent network of gene regulation and
metabolism.

Here we introduce the concept of network response to
localized perturbations analysis. This analysis will reveal that
perturbations in gene regulation affect the system in a dramatically
different way than perturbations in metabolism.
Thus we study the response to \textit{localized} perturbations. We
denote such perturbations by $\mbox{Per}(X, q)$, where $X$ is the
domain, in which the perturbation is localized ($X\in \{R, I, M,
T\}$, with $T$ representing the total network $G$, i.e., the case of
non-localized perturbations). The perturbation size $q=1-p$ is
measured in fractions of the total network size $N = |G|$. A perturbation
$\mbox{Per}(M, 0.1)$ thus is a perturbation in the metabolic domain
with (on average) $0.1 |G| $ nodes initially affected. Note that
sizes $q$ of such localized perturbations are limited by the domain
sizes, e.g., $q |G| < |G_R|$ for a perturbation in the gene
regulatory domain.

After the initial removal of a fraction $q$ of the vertices from the
domain $X$ the stepwise dynamics described above will lead to the
deactivation of further nodes and run into a frozen state
$\vec{\sigma}_\infty$ in which only a fraction $A(X,q)$ of the
vertices are unaffected by the perturbation (i.e., are still in state
$1$). In addition to being directly affected by failing neighbors,
in the process of network fragmentation nodes may also become parts
of small components disconnected from the network's core, and could
in this sense be considered non-functional; we therefore also monitor
the relative sizes of the largest (weakly) connected component in the
frozen state, $B(X,q)$, for different initial perturbation sizes.
 
In the limit of infinite system size we could expect a direct
investigation of $B(X,q)$ as a function of $q$ to yield the critical
perturbation size $q_c=1-p_c$ at which $B$ vanishes. In our finite system,
however, we have to estimate $q_c$; following Radicchi and Castellano
\cite{PhysRevE.93.030302} and Radicchi
\cite{radicchi_predicting_2015} we measure the fluctuation of
$B(X,q)$ which serves as our order parameter and look for the peak
position of the susceptibility
\begin{align}
  \tilde{\chi}(X,q) = \langle B_\infty^2 \rangle - \langle B_\infty
  \rangle^2
  \label{eq:susceptibility}
\end{align}
as a function of parameter $q$ in order to estimate the transition
point from the finite system data.  Supplementary
Figure~\ref{fig:overview} schematically illustrates the
analysis.

\section{Randomization Schemes\label{sec:randomize}}
In order to interpret the actual responses of a given network 
to perturbations, one usually contrasts them to those of suitably 
randomized versions of the network at hand. Thereby, the 
often dominant effect of the node degree distribution of a 
network can be accounted for and the effects of higher-order 
topological features that shape the response of the network 
to perturbations can be studied systematically. 

The same is true for the localized perturbation response analysis
introduced here. In fact, due to the substantially larger number of
links from gene regulation to metabolism (both, directly and via the
interface component of the interdependent network) than from
metabolism to gene regulation we can already expect the response to
such localized perturbations to vary.

Here we employ a sequence of ever more stringent randomization
schemes to generate sets of randomized networks serving as null
models for the localized perturbation response analysis. In all of
the four schemes the edge-switching procedure introduced by Maslov
and Sneppen \cite{Maslov910} is employed which conserves the in- and
out-degrees of all vertices.

Our most flexible randomization scheme (DOMAIN) only considers the
domains of the source and target vertices of an edge (SD and TD):
only pairs of edges are flipped which share both, the source and the
target domain (e.g., both link a vertex in the metabolic domain to a
vertex in the interface).  The remaining three randomization schemes
all add an additional constraint. The DOMAIN\_LCE randomization
further requires the edges to be of the same logical categories of an
edge (i.e., C, D, or R), while the DOMAIN\_BCV scheme only switches
edges whose target vertices also share the same biological category
of a vertex, BCV.  The strictest randomization, DOMAIN\_BCE, finally,
only considers edges with, additionally, the same biological category
of an edge, BCE. A tabular overview of the four schemes is given in
Supplementary Table \ref{tab:randschemes}.

\section{Results\label{sec:results}}
The main feature of our reconstructed network, the three-domain
structure based on the biological role of its constituents, allows us 
to study the influence of localizing the initial perturbation. Thus,
although we will not focus on (topological) details of the graph here
(which will be presented elsewhere \cite{grimbs_integrated_2016}),
already from the vertex and edge counts in Figure~\ref{fig:sketch} we
see that the domains are of different structure. While the regulatory 
and the metabolic subgraphs, $G_R$ and $G_M$ have average (internal)
degrees of about $1.9$ and $2.5$, the interface subgraph, $G_I$, is
very sparse with $\langle k \rangle\approx 0.6$ and we can expect it
to be fragmented. Hence, in the following we decide to only perturb
in $G_R$ and $G_M$.

In a first step we sample some full cascade trajectories in order to
check our expectation of different responses of the system to small
perturbations applied in either $G_R$ or $G_M$; two rather large
values of $q$ are chosen and the raw number of unaffected nodes is
logged during the cascade.  Indeed, already this first approach
implies a different robustness of the gene regulatory and the
metabolic domains in terms of the transmission of perturbation
cascades to the other domains. Cascades seeded in the metabolic
domain of the network tend to be rather restricted to this domain,
while the system seems much more susceptible to small perturbations
applied in the gene regulatory domain.  This effect can be seen both
in the overall sizes of the aggregated cascades as well as in the
domain which shows the largest change with respect to the previous
time step, which we indicate by black markers in
Figure~\ref{fig:states}.  More sample trajectories are shown in the
Supplementary Materials and, although they illustrate occasionally
large fluctuations between the behaviors of single trajectories, they
are consistent with this first observation. They also show that
considerably larger metabolic perturbations are needed for large
cascades and back-and-forth propagation between domains to emerge.

\begin{figure*}[h]
  \centering
  \includegraphics[width=.95\textwidth]
                  {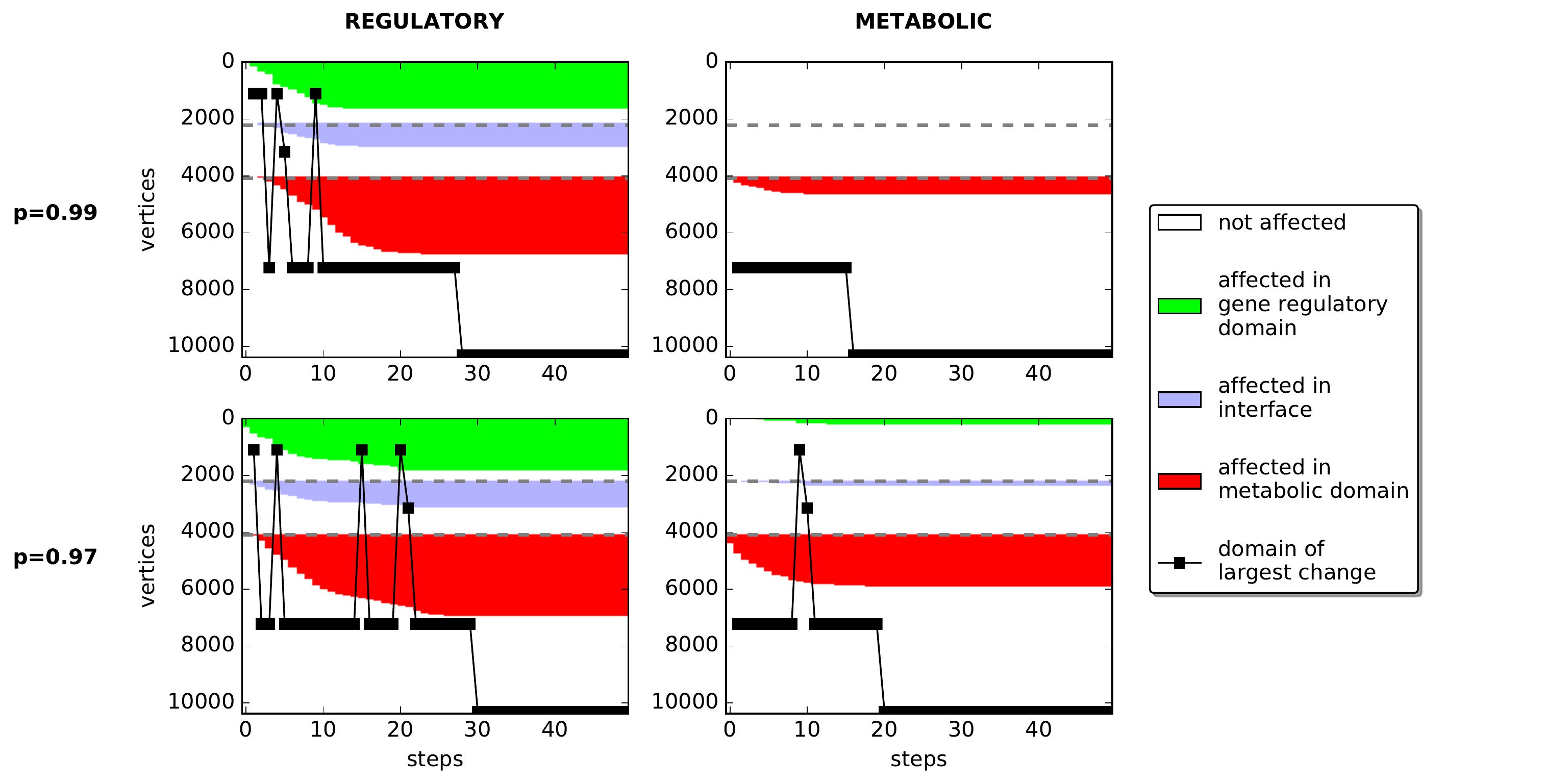}
  \caption{\label{fig:states} Sample trajectories of the integrative
    \textit{E.~coli} network after perturbations of size $q=1-p=0.01$
    (top row) and $q=0.03$ (bottom row). The left column illustrates
    perturbations seeded in the regulatory domain, while the right
    column shows results for perturbations in the metabolic domain.}
\end{figure*}

After this first glance at the system we aim for a more systematic
approach and apply our analysis as described above: we compute
cascade steady-states $\vec{\sigma}_\infty$ but now we choose the
largest (weakly) connected component $B(X,q)$ as the order parameter
and compute the susceptibility according to equation 
$(\ref{eq:susceptibility})$, the peak-position of which, when
considered as a function of $q=1-p$, we use as a proxy for the
perturbation size at which the interdependent system breaks down.

\begin{figure}[h]
  \includegraphics[width=.95\columnwidth]
                  {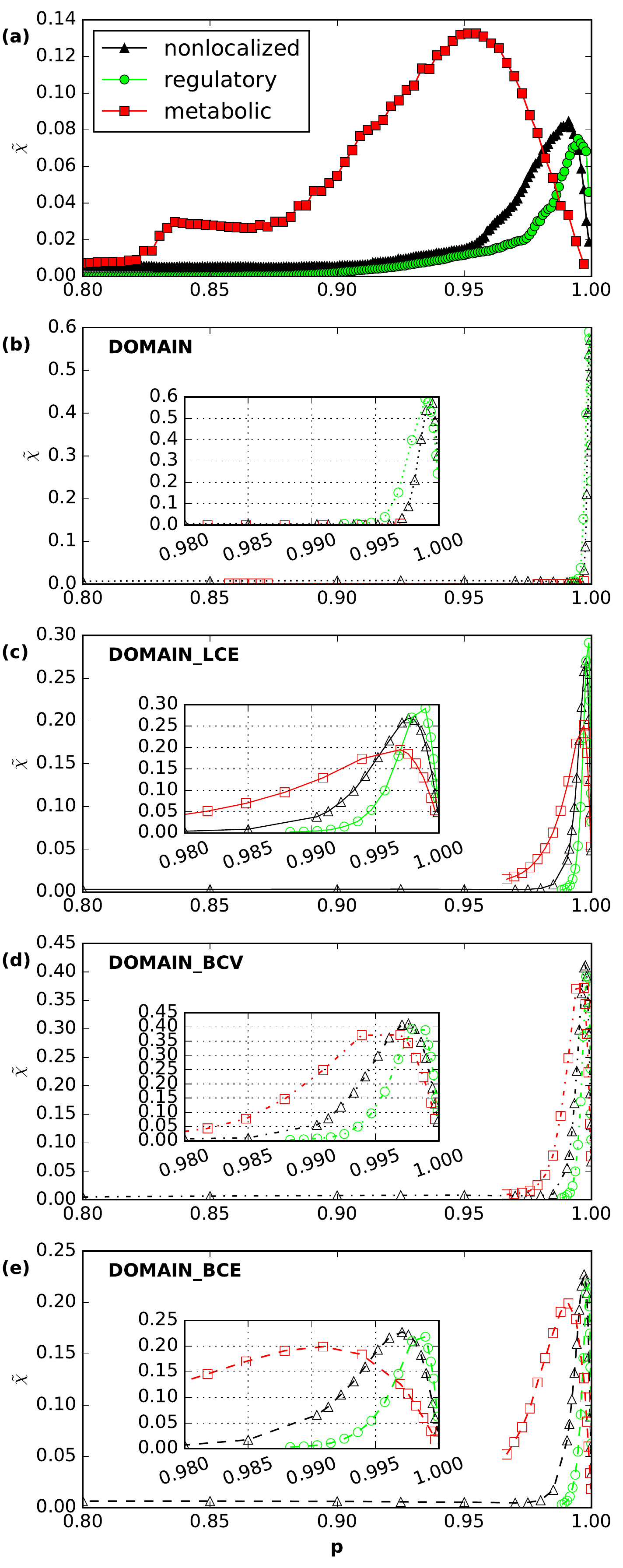}
  \caption{\label{fig:suscept} The susceptibility of the integrative
    \textit{E.~coli} model and its randomized versions for
    perturbations in different domains. The results for the original
    graph are shown in the \textit{top} frame, while the subsequent
    frames show results for the four different randomization schemes
    with the least strict on top and the strictest one at the
    bottom. The original system is more robust than its randomized
    versions; perturbations in the metabolism consistently need to be
    larger than in the regulatory part in order to reach the maximum
    susceptibility.}
\end{figure}

The results for different initially perturbed domains illustrate
that, indeed, a considerably lower $p_c$ (i.e., larger critical
perturbation size $q_c$) is estimated in the case of metabolic
perturbations compared to regulatory or non-localized ones
(Figure~\ref{fig:suscept}, panel a). For each point we average $500$
runs for the corresponding set of parameters.

In order to assess whether the above-described behavior is due to
specific properties of the network we use the sets of randomized
graphs. For each of the four randomization schemes we prepared $500$
graph instances and repeated the analysis for each of them as done
before for the single original graph. The corresponding results for
the susceptibility (Figure~\ref{fig:suscept}, panels b--e)
yield two major observations: firstly, metabolic perturbations still
lead to, albeit only slightly, higher $q_c=1-p_c$ estimates (with
exception of DOMAIN randomization). Thus, the system's tendency to be
more robust towards metabolic perturbations is largely
preserved. Secondly, we see that overall the original network seems
to be much more robust than the randomized networks; very small
perturbations are sufficient to break the latter ones. The robustness
of the original graph, thus, cannot be solely due to the edge and
vertex properties kept in the randomization schemes.

\begin{figure}[h]
  \centering
  \includegraphics[width=\columnwidth]
                  {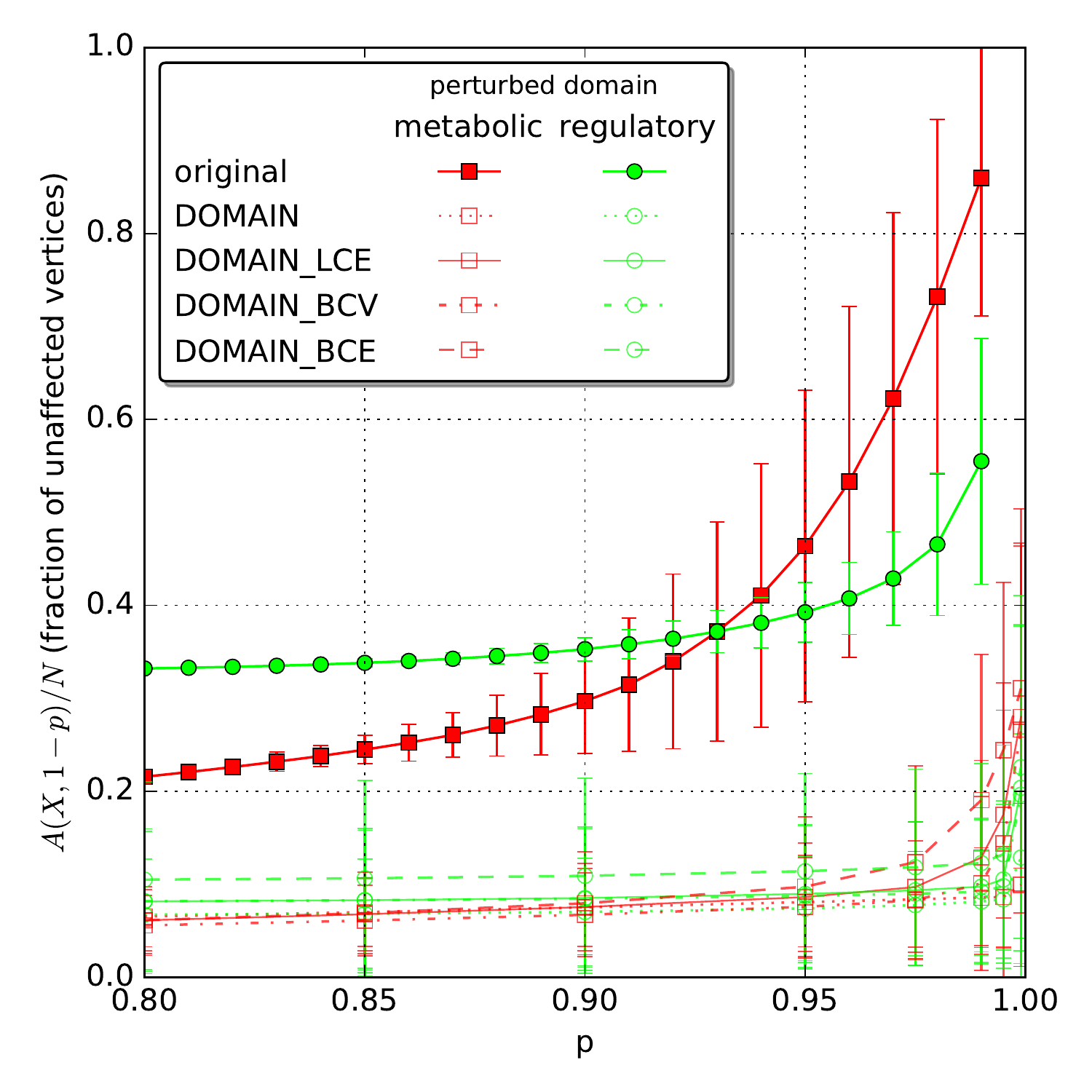}
  \caption{\label{fig:frac} Mean fractions (averaged over $5,000$
    runs) of unaffected vertices after initial perturbations of
    different sizes seeded in the gene regulatory (filled green dots)
    and in the metabolic domain (filled red squares). The curves at
    the bottom show the averaged results over $500$ shuffles for each
    randomization scheme.}
\end{figure}

Finally, let us focus on the practical aspect of these findings.
Beyond the careful statistical analysis described above, a quantity
of practical relevance is the average size of the unaffected part of
the system under a perturbation. For this purpose, we examine the
fractions of unaffected vertices, $A(X,q)$, after cascades emanating
from perturbations of different sizes and seeded in different
domains, regardless of the resulting component structure and for
both, the original graph and the shuffled ones
(Figure~\ref{fig:frac}).

The number of unaffected vertices for the real network is much larger
than for all four randomization schemes, suggesting a strong overall
robustness of the biological system. Distinguishing, however, between
the metabolic and the gene regulatory components reveals that the
metabolic part is substantially more robust than the regulatory part
(for not too large initial perturbations, $p>0.94$).

\section{Discussion and Outlook \label{sec:conclusion}}

We investigated the spreading of perturbations through the
three domains of a graph representation of the integrated system of
\textit{E.~coli}'s gene regulation and metabolism. Our results quantify 
the resulting cascading failures as a function of size and
localization of the initial perturbation.

Our findings show that the interdependent network of gene regulation 
and metabolism unites sensitivity and robustness by showing different 
magnitudes of damage dependent on the site of perturbation. 

While the interdependent network of these two domains is in general
much more robust than its randomized variants (retaining domain
structure, degree sequence, and major biological aspects of the
original system), a pronounced difference between the gene regulatory
and metabolic domain is found: Small perturbations originating
in the gene regulatory domain typically trigger far-reaching
system-wide cascades, while small perturbations in the metabolic
domain tend to remain more local and trigger much smaller cascades of
perturbations.

In order to arrive at a more mechanistic understanding of this
statistical observation, we estimated the percolation threshold of
the system, $p_c$, and found that it is much lower (i.e, larger
perturbations, $q_c=1-p_c$, are required) for perturbations seeded in
the metabolic domain than for those applied to the gene regulatory
domain.

This is in accordance with the intuition that the metabolic
system is more directly coupled to the environment (via the uptake
and secretion of metabolic compounds) than the gene regulatory
domain. The distinct perturbation thresholds therefore allow for 
implementing a functionally relevant balance between robustness 
and sensitivity: The biological system can achieve a robustness 
towards environmental changes, while -- via the more sensitive gene
regulatory domain -- it still reacts flexibly to other systemic
perturbations. 

Discovering this design principle of the biological system required 
establishing a novel method of analyzing the robustness of 
interdependent networks, the network response to localized perturbations: 
An interdependent network can have markedly different percolation 
thresholds, when probed with perturbations localized in one network 
component compared to another. 

Lastly, we would like to emphasize that the application of 
the theoretical concepts of interdependent networks to 
real-life systems involves several non-trivial decisions: 

In the vast majority of (theoretical) investigations, 
two definitions of interdependent networks coincide: 
the one derived from a distinction between dependency links 
and connectivity-representing links and the one based 
on two functionally distinguishable, but interconnected subnetworks. 

Here we have three classes of nodes: those involved in gene regulation, 
metabolic nodes, and nodes associated with the (protein) interface 
between these two main domains. These nodes are interconnected 
with (functionally) different classes of links. 
These link classes are necessary to define meaningful update rules 
for perturbations.  As a consequence, the notion of dependency links 
vs.~connectivity links is no longer applicable. We expect that such 
adjustments of the conceptual framework will often be required 
when applying the notion of interdependent networks to real-life systems.

As mentioned above, one major task when dealing with biological data
is to abstract from the minor but keep the essential details; we have
outlined that in this study we chose to keep a rather high level of
detail.

With only incomplete information available, a challenge is to find 
the right balance between radical simplifications of systemic descriptions 
and an appropriate level of detail still allowing for a meaningful 
evaluation of biological information. Here we incorporate high level 
of detail in the structural description, distinguishing between 
a comparatively large number of node and link types. This rich 
structural description, together with a set of update rules 
motivated by general biological knowledge, allows us to assess 
the dynamical/functional level with the comparatively simple 
methods derived from percolation theory. 

An important question is, whether the analysis of the fragmentation
of such a network under random removal of nodes can provide a
reliable assessment of functional properties, since the response of
such a molecular network clearly follows far more intricate dynamical
rules than the percolation of perturbations can suggest.

A future step could include the construction of a Boolean network
model for the full transcriptional regulatory network and the
connection of this model to flux predictions obtained via flux
balance analysis, a first attempt of which is given in Samal and Jain
\cite{samal_regulatory_2008} (where the model of Covert \textit{et
  al.~}\cite{covert_integrating_2004} with still fewer
interdependence links has been used).

Our perturbation spreading approach might help bridging the gap
between theoretical concepts from statistical physics and biological
data integration: Integrating diverse biological information into
networks, estimating 'response patterns' to systemic perturbations
and understanding the multiple systemic manifestations of perturbed,
pathological states is perceived as the main challenge in systems
medicine (see, e.g., Bauer \textit{et
  al.~}\cite{bauer2016interdisciplinary}). Concepts from statistical
physics of complex networks may be of enormous importance for this
line of research \cite{Barabasi:2011iha,huett2014}.

While the simulation of the full dynamics is still problematic as our  
knowledge of the networks is still incomplete, our present strategy 
extracts first dynamical properties of the interdependent networks. 
At a later time point, we can expect qualitatively advances from 
full dynamical simulations, however, dependent on the quality of the 
data sets. 

On the theoretical side, future studies might shift the focus onto 
recasting the system into
an appropriate spreading model, e.g., in the form of an unordered
binary avalanche \cite{samuelsson_exhaustive_2006,gleeson_mean_2008},
or as an instance of the Linear Threshold model
\cite{watts_simple_2002} with a set of links with a very high and a
second set with a very low transmission probability (C/R and D-links,
respectively).

Radicchi \cite{radicchi_percolation_2015} presents an approach for
the investigation of the percolation properties of finite size
interdependent networks with a specific adjacency matrix with the
goal of loosening some of the assumptions underlying the usual models
(e.g., infinite system limit, graphs as instances of network
model). While this formalism allows for the investigation of many
real-world systems there are still restrictions as to the possible
level of detail. In our special case, for instance, a considerable
amount of information would be lost if the system was restricted to
vertices with connections in both the C/R- and D-layers.

The existence of different percolation thresholds for 
localized perturbations in interdependent networks may reveal itself 
as a universal principle for balancing sensitivity and robustness 
in complex systems. The application of these concepts to a wide range 
of real-life systems is required to make progress in this direction.

\begin{acknowledgments}
S.B.~and M.H.~acknowledge the support of Deutsche
Forschungsgemeinschaft (DFG), grants BO 1242/6 and HU 937/9.
\end{acknowledgments}

\section*{Author Contributions}
M.H.~and S.B.~designed and supervised the study; D.K.~and
A.G.~performed the reconstruction, simulations and analyses; and
D.K., A.G., S.B.~ and M.H.~wrote the manuscript.

\section*{Competing Financial Interests}
The authors declare no competing financial interests.



\onecolumngrid
\clearpage

\begin{center}
  \textbf{Supplementary Material to\\\large The interdependent
    network of gene regulation and metabolism is robust where it
    needs to be}
\end{center}

\setcounter{page}{1}
\setcounter{section}{0}

\renewcommand{\thefigure}{S\arabic{figure}} \setcounter{figure}{0}
\renewcommand{\thetable}{S\arabic{table}} \setcounter{table}{0}

\renewcommand{\figurename}{FIG.}
\renewcommand{\tablename}{TABLE}

\renewcommand{\textfraction}{0.01}
\setlength{\parindent}{0pt}

Here, we provide supplementary information to the main manuscript:
\begin{enumerate}[label=\bfseries\Roman*.]
 \item A description of the analysis including a schematic overview,
 \item An explanation of the notation of vertex and edge categories,
 \item Tables summarizing the four custom-built randomization
   schemes, the quantities and parameters shown/used in the
   particular figures as well as for the introduced vertex and edge
   categories
 \item A collection of sample trajectories, $\{V_t(X,q)\}_t$, to be
   compared to Fig.~\ref{fig:states} in the main manuscript.
\end{enumerate}

\section{Description and Schematic Overview of the analysis}
\begin{figure}[htb]
 \centering
 \begin{tikzpicture}
  \node (T1) {\begin{minipage}{4.7em}\raggedright original network
      $G$\end{minipage}};
  \node[Square,right=of T1] (P1) {$\mathrm{Per}\left(X,q\right)$};
  \node[right=7em of P1] (T2) {statistical evaluation:};
  \node[below=0.5em of T2.west,anchor=north west] (T3)
       {\begin{minipage}{13em}
           \begin{itemize}
           \item susceptibility $\chi\left(X,q\right)$ \\[-1.75em]
           \item variance $var\left[B\left(X,q\right)\right]$
           \end{itemize}
       \end{minipage}};
  \node[rhombus,fill=black!15,below=7em of P1.east] (D)
       {\begin{minipage}{6em}\centering run perturbation
           dynamics\end{minipage}};
  \node[square,below left=2em of D] (B1)
       {\begin{minipage}{8em}\raggedright select domain $X$ and
           perturbation size
           $q$\\\vphantom{$V_{0}\left(X,q\right)$}\end{minipage}};
  \node[anchor=south] (T4) at (B1.south) {$V_{0}\left(X,q\right)$};
  \coordinate[left=1.75em of T4] (T4a);
  \node[square,right=2em of D] (B2) {$\left\lbrace
    V_{t}\left(X,q\right) \right\rbrace_{t}$};
  \node[square,below=1.5em of B2] (B3)
       {\begin{minipage}{11.5em}\raggedright size of potentially
           unaffected network $A\left(X,q\right) = \left| V \setminus
           V_{\infty}\left(X,q\right)\right|$\end{minipage}};
       \coordinate[right=2.75em of B3] (B3a);    
  \node[square,below=1.5em of B3] (B4)
       {\begin{minipage}{13em}\raggedright size of the largest
           connected component of the unaffected network,
           $B\left(X,q\right)$\end{minipage}}; \coordinate[right=2em
         of B4] (B4a); \path let \p1=(B1.west),\p2=(B4.south) in
       coordinate (P2c) at (\x1,\y2);
  \node[draw,rectangle,anchor=south west] (P2) at (P2c)
       {$\mathrm{Per}\left(X,q\right)$};
  \begin{pgfonlayer}{background}
   \node[Box,fit=(D)(B1)(B4)] (BOX) {};
  \end{pgfonlayer}
  \path let \p1=(T1.west),\p2=(BOX.south) in coordinate (T5c) at
  (\x1,\y2);
  \node[below=of T5c,anchor=west] (T5)
       {\begin{minipage}{6.5em}\raggedright randomization scheme
           $i$\end{minipage}};
  \node[right=2em of T5] (T6) {\begin{minipage}{8.1em}\raggedright
      set of randomized networks $\left\lbrace G^{\left(R,i\right)}
      \right\rbrace$\end{minipage}};
  \node[Square,right=2em of T6] (P2) {$\mathrm{Per}\left(X,q\right)$};
  \node[right=4.5em of P2] (T7) {statistical evaluation:};
  \node[below=0.5em of T7.west,anchor=north west] (T8)
       {\begin{minipage}{15em}
           \begin{itemize}\item susceptibilities
             $\chi^{\left(R,i\right)}\left(X,q\right)$ \\[-1.75em]
           \item variances $var\left[B^{\left(R,i\right)}\left(X,q\right)\right]$ \end{itemize}\end{minipage}};
  
  \node[above=1em of T1.west,blue] (L1) {\textbf{(a)}};
  \path let \p1=(L1),\p2=(BOX.north) in coordinate (L2c) at (\x1,\y2);
  \node[above=0em of L2c.west,blue] {\textbf{(b)}};
  \node[above=1em of T5.west,blue] {\textbf{(c)}};
  
  \draw[thick] (P1.south west) to (BOX.north west);
  \draw[thick] (P1.south east) to (BOX.north east);
  \draw[-latex,thick] (T4a) to (T4);
  
  \draw[-latex,thick] (T1) to (P1);
  \draw[-latex,thick] (P1) to node[above] {\footnotesize multiple runs} (T2);
  \draw[-latex,thick] ($(T1.west)+(0,-2.5)$) to ($(T1.west)+(1,-2.5)$) to (D.west);
  \draw[-latex,thick] (B1.north) to (D.west);
  \draw[-latex,thick] (D.east) to (B2);
  \draw[-latex,thick] (B2) to (B3);
  \draw[-latex,thick] (B3) to (B4);
  \draw[-latex,thick] (B3) to (B3a); 
  \draw[-latex,thick] (B4) to (B4a);
  \draw[-latex,thick] (T5) to (T6);
  \draw[-latex,thick] (T6) to (P2);
  \draw[-latex,thick] (P2) to node[above] {\begin{minipage}{3.3em}\footnotesize multiple runs\end{minipage}} (T7);
 \end{tikzpicture}
 \caption{Schematic representation of the network response to
   localized perturbations analysis: (a) The graph $G$ enters the
   perturbation algorithm $\mbox{Per}(X, q)$ characterized by the
   domain $X$ and the size of the perturbation $q$; statistics over
   multiple runs are then evaluated in terms of susceptibilities and
   variances; (b) details of the perturbation algorithm; based on $X$
   and $q$ the set $V_0(X, q)$ of initially perturbed nodes is
   randomly selected; running the perturbation dynamics leads to a
   time series of affected nodes, $\{V_t(X, q)\}_t$, which is
   subsequently evaluated yielding the size of the unaffected network
   (which can be derived from the number of nodes in the set $V_t(X,
   q)$ in the asymptotic regime) and the relative size of the largest
   (weakly) connected component of the unaffected network, $B(X, q)$;
   (c) same as (a), but for randomized networks.}\label{fig:overview}
\end{figure}
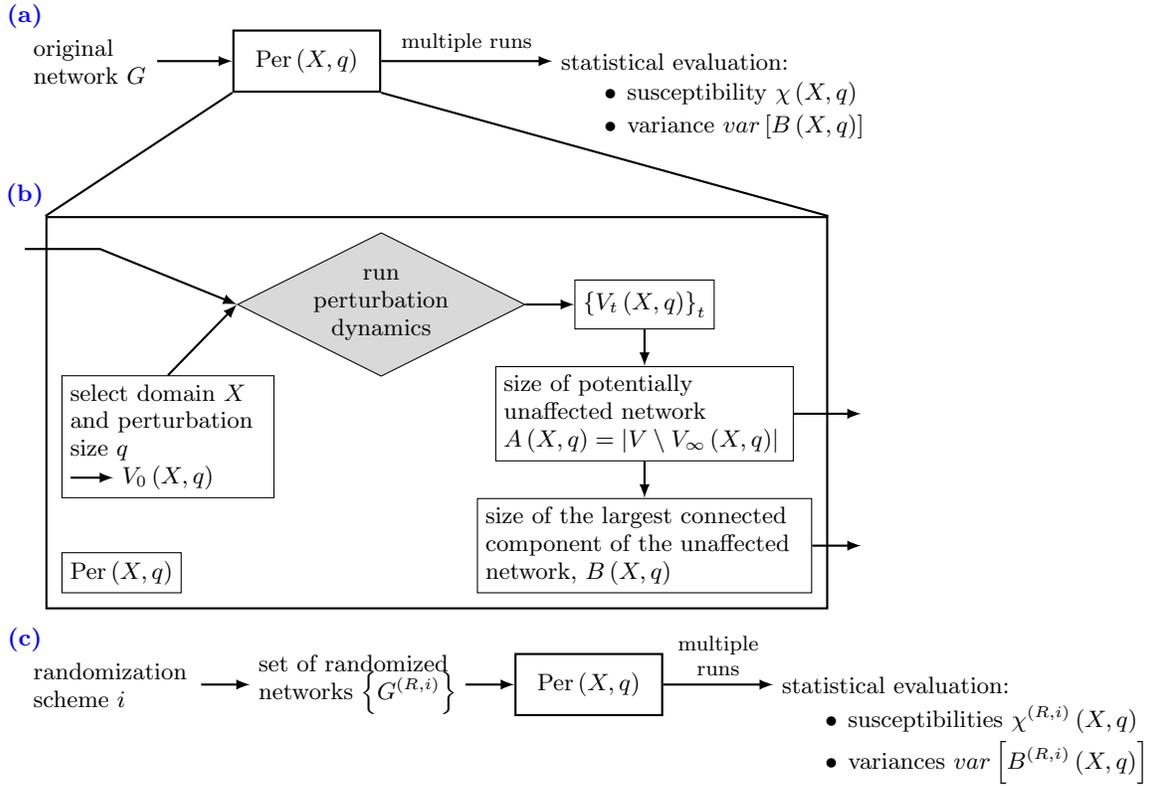

The network response to localized perturbations analysis presented
here (see Figure~\ref{fig:overview}), is a multi-step method
entailing several runs of the perturbation algorithm, $\mbox{Per}(X,
q)$, and the statistical evaluation of these runs. More precisely,
for each analysis 500 runs of the perturbation algorithm have been
performed and the set of the remaining vertices have been evalueted,
e.g.,~in terms of the susceptibility, $\chi\left(X,q\right)$ (part a
in Figure~\ref{fig:overview}).\par For a single run of the
perturbation algorithm, $\mbox{Per}(X, q)$, a domain $X$ has to be
chosen where a perturbation of size $q=1-p$ (measured as a fraction
of vertices) will be applied,
 \begin{align*}
  X &\in \left\lbrace R, I, M, T \right\rbrace
  \intertext{with $R$ -- regulatory domain, $I$ -- protein interface,
    $M$ -- metabolic domain, $T$ -- total network.}
 \end{align*}
 Based on $X$ and $q$ the set of initially perturbed vertices,
 $V_{0}\left(X, q\right)$, is randomly selected.  The state of the
 system can also be described as a vector of Boolean state variables,
 $\vec{\sigma}$,
 \begin{align*}
   \sigma_{i} \in \left\lbrace 0,1 \right\rbrace, \; i =
   \left\lbrace1,\ldots,\left|G\right|\right\rbrace
 \end{align*}
 where 0 denotes a perturbed vertex and 1 an unaffected one.\par
 Running the perturbation dynamics described in the main manuscript
 will (probably) cause the failure of further vertices resulting in a
 time series of affected vertices, $\{V_t(X, q)\}_t$ (or,
 equivalently, to the trajectory $\sigma(t,\,X,q)$). The size of the
 affected network after $t$ propagation steps can be described as
 \begin{align*} 
  \left| V_{t}\left(X,q\right) \right| &= \left|G\right| -
  \sum\limits_{i=1}^{\left|G\right|} \sigma_{i} \left(t,\, X,q
  \right)
 \end{align*} 
 From the set of affected vertices in the in the asymptotic regime,
 $V_{\infty}\left(X,q\right)$, the size of the 
 unaffected network, $A\left(X,q\right)$, and the size of the largest
 (weakly) connected component ($LCC$) of the unaffected network,
 $B\left(X,q\right)$ are computed (Figure~\ref{fig:overview}, part
 b),
 \begin{align*}
 A\left(X,q\right) &= \left| V \setminus V_{\infty}\left(X,q\right)
 \right|\text{,} \\
 B\left(X,q\right) &= \left| \mathbf{LCC}\left[V \setminus
 V_{\infty}\left(X,q\right)\right] \right|\text{.}
 \end{align*}
Randomized networks (we used sets of $500$ instances for each of the
randomization schemes) can be passed to the algorithm instead
(Figure~\ref{fig:overview}, part c).

\FloatBarrier
\newpage
\section{Notation of vertex and edge categories}
The following notation concerning vertices $v_i \in V$ and edges $e_i
\in E$ and their properties have been used.
In our integrated network model a vertex is characterized by its
biological function yielding seven unique \textit{biological
  categories of a vertex} (BCVs), 'reaction' ($rxn$), 'compound'
($cmp$), 'gene' ($gn$), 'protein monomer' ($pm$), 'protein-protein
complex' ($ppc$), 'protein-compound complex' ($pcc$), and
'protein-rna complex' ($prc$; see Table~\ref{tab:BCVs}),
\begin{displaymath}
v_i^{\text{BCV}} \in \{rxn, cmp, gn, pm, ppc, pcc, prc\}.
\end{displaymath}
Introducing an additional vertex classification facilitates the
assignment to one of the three functional domains as well as the edge
characterization. The \textit{domain-related categories of a vertex}
(DCVs) are eight-fold: 'gene' ($g$), 'protein' ($p$), 'complex' ($x$),
'enzyme' ($z$), 'reaction' ($r$), 'compound' ($c$), 'educt' ($e$),
'product' ($d$),
\begin{displaymath}
v_i^{\text{DCV}} \in \{g, p, x, z, r, c, e, d\}.
\end{displaymath}
While the categories $g$ and $r$ are a one-to-one translations of the
corresponding BCVs, i.e., $gn$ and $rxn$, the domain-related category
$c$ only comprises vertices of BCV $cmp$ but the inverse does not
hold. The remaining five categories are ambiguous assignments for
both BCVs to DCVs and vice versa. The complete mapping of BCVs onto
DCVs is given in Table \ref{tab:DCVs}.\par
An edge is characterized by its source and target vertices,
$v_s^{(i)}$ and $v_t^{(i)}$, and their corresponding domains ('source
domain', SD and 'target domain', TD), as well as the edge's
\textit{logical category} and its \textit{biological category}; an
edge is given by
\begin{displaymath}
e_i = (v_s^{(i)}, v_t^{(i)}) \, , \; v_j^{(i)} \in \{G_R, G_I, G_M \},
\end{displaymath}
thus determining SD and TD. The \textit{logical category of an edge}
(LCE) determines, qualitatively speaking, whether a perturbation will
propagate along this edge via a logical \texttt{AND} or a logical
\texttt{OR}; the three categories are 'conjunct' ($C$), 'disjunct'
($D$) and 'regulation' ($R$),
\begin{displaymath}
  e_i^\text{LCE} \in  \{C, D, R \}.
\end{displaymath}
As an illustration of the potential linkages, two case examples are
presented in Figure \ref{fig:BiologicalCaseExamples}.
\begin{figure}[htb]
 \centering
 \begin{tikzpicture}
  \node[reaction,fill=mygreen,scale=1.5,label={[yshift=-0.5em]below:{EC
        6.2.1.5}}] (E) {};
  \node[protein,fill=mygray,scale=1.5,above=2.5em of
    E.north,label={left:{$\left[\text{SucC}\right]_{2}\!\left[\text{SucD}\right]_{2}$}}]
  (CD) {};
  \node[protein,fill=myred,scale=1.5,above left=2em and 1.5em of
    CD.north,label={left:{SucC}}] (C) {};
  \node[protein,fill=myred,scale=1.5,above right=2em and 1.5em of
    CD.north,label={right:{SucD}}] (D) {};
  \node[gene,fill=myyellow,scale=1.5,above=1.5em of
    C.north,label={left:{sucC}}] (c) {};
  \node[gene,fill=myyellow,scale=1.5,above=1.5em of
    D.north,label={right:{sucD}}] (d) {};
  \node[compound,fill=myblue,scale=1.5,above left=0.5em and 3em of
    E,label={left:{Succ}}] (Succ) {};
  \node[compound,fill=myblue,scale=1.5,left=3em of
    E,label={left:{CoA}}] (CoA) {};
  \node[compound,fill=myblue,scale=1.5,below left=0.5em and 2.75em of
    E,label={left:{ATP}}] (ATP) {};
  \node[compound,fill=myblue,scale=1.5,above right=0.5em and 3em of
    E,label={right:{S-CoA}}] (SuccCoA) {};
  \node[compound,fill=myblue,scale=1.5,right=3em of
    E,label={right:{ADP}}] (ADP) {};
  \node[compound,fill=myblue,scale=1.5,below right=0.5em and 2.75em
    of E,label={right:{Pi}}] (Pi) {};
  \draw[encodes] (c) to node[right,pos=0.7] {\scriptsize D} (C);
  \draw[encodes] (C.south) to node[left,pos=0.2] {\tiny 2}
  node[above,pos=0.8] {\scriptsize C} (CD.north);
  \draw[encodes] (d) to node[left,pos=0.7] {\scriptsize D} (D);
  \draw[encodes] (D.south) to node[right,pos=0.2] {\tiny 2}
  node[above,pos=0.8] {\scriptsize C} (CD.north);
  \draw[encodes] (CD) to node[right,pos=0.75] {\scriptsize D} (E);
  \draw[encodes] (Succ.east) to node[above right,yshift=-0.5em]
       {\scriptsize C} (E);
  \draw[encodes] (CoA.east) to node[above right,yshift=-0.2em]
       {\scriptsize C} (E);
  \draw[encodes] (ATP.east) to node[above right,xshift=-0.2em]
       {\scriptsize C} (E);
  \draw[encodes] (E) to node[above right,xshift=-0.2em] {\scriptsize
    D} (SuccCoA.west);
  \draw[encodes] (E) to node[above right,yshift=-0.2em] {\scriptsize
    D} (ADP.west);
  \draw[encodes] (E) to node[above right,yshift=-0.5em] {\scriptsize
    D} (Pi.west);
  \draw[activates] (ADP.30) to[out=5,in=0,looseness=3.5]
  node[above,pos=0.96] {\scriptsize R} (CD.43);
  \node[compound,fill=myblue,scale=1.5,right=17em of
    CD,label={left:{chorismate}}] (Ch) {};
  \node[reaction,fill=mygreen,scale=1.5,above right=0.5em and 3em of
    Ch,label={[yshift=-0.3em]below:{EC 5.4.4.2}}] (E1) {};
  \node[compound,fill=myblue,scale=1.5,below right=0.5em and 3em of
    E1,label={right:{isochorismate}}] (I) {};
  \node[reaction,fill=mygreen,scale=1.5,below right=0.5em and 3em of
    Ch] (E2) {};
    
  \node[protein,fill=mygray,scale=1.5,above=2.5em of
    E1.north,label={left:{$\left[\text{MenF}\right]_{2}$}}] (X) {};
  \node[protein,fill=myred,scale=1.5,above=1.5em of
    X.north,label={left:{MenF}}] (MenF) {};
  \node[gene,fill=myyellow,scale=1.5,above=1.5em of
    MenF.north,label={left:{menF}}] (menF) {};
  \node[protein,fill=myred,scale=1.5,below=2em of
    E2.south,label={left:{EntC}}] (EntC) {};
  \node[gene,fill=myyellow,scale=1.5,below=1.5em of
    EntC,label={left:{entC}}] (entC) {};
  \draw[encodes] (menF) to node[right,pos=0.7] {\scriptsize D}
  (MenF); \draw[encodes] (MenF) to node[right,pos=0.25] {\tiny 2}
  node[right,pos=0.8] {\scriptsize C} (X);
  \draw[encodes] (X) to node[right,pos=0.75] {\scriptsize D} (E1);
  \draw[encodes] (Ch.east) to node[above,pos=0.7] {\scriptsize C} (E1);
  \draw[encodes] (E1) to node[above,pos=0.8] {\scriptsize D} (I.west);
  \draw[encodes] (entC) to node[right,pos=0.7] {\scriptsize D} (EntC);
  \draw[encodes] (EntC) to node[right,pos=0.75] {\scriptsize D} (E2);
  \draw[encodes] (Ch.east) to node[below,pos=0.7] {\scriptsize C} (E2);
  \draw[encodes] (E2) to node[below,pos=0.8] {\scriptsize D} (I.west);
  
  \path let \p1=(CoA.west),\p2=(menF.north) in coordinate (L1) at
  (\x1,\y2);
  \node[left=of L1] {\textbf{A}};
  \path let \p1=(Ch.west),\p2=(menF.north) in coordinate (L2) at
  (\x1,\y2);
  \node[left=6em of L2] {\textbf{B}};
 \end{tikzpicture}
 \caption{Two biological case examples of potential linkages,
   \textbf{A} succinyl-CoA synthetase (EC 6.2.1.5) and \textbf{B}
   isochorismate synthase (EC 5.4.4.2). The vertices are denoted by
   the common biological abbreviation (see EcoCyc webpage) and the
   respective biological category of a vertex (BCV):
   \protect\tikz\protect\node[gene,fill=myyellow,scale=0.7]{};~gene,
   \protect\tikz\protect\node[protein,fill=myred,scale=0.7]{};~protein
   monomer,
   \protect\tikz\protect\node[protein,fill=mygray,scale=0.7]{};~protein-protein-complex,
   \protect\tikz\protect\node[reaction,fill=mygreen,scale=0.7]{};~reaction,
   \protect\tikz\protect\node[compound,fill=myblue,scale=0.7]{};~compound. Moreover,
   the involved edges are labeled with the corresponding logical
   category of an edge (LCE): 'conjunct' ($C$), 'disjunct' ($D$) and
   'regulation' ($R$). The additional numbers indicate stoichiometric
   coefficients for the complex
   formation.}\label{fig:BiologicalCaseExamples}
\end{figure}

The biological categories of an edge (BCEs) are derived from
combinations of the domain-related categories of a vertex (DCVs), plus
'transport' ($t$) and regulation ($r^*$),
\begin{displaymath}
e^\text{BCE}_i \in \{g\rightarrow p, p\rightarrow x, c\rightarrow x,
z\rightarrow r, e\rightarrow r, r\rightarrow d, t, r^* \}\text{.}
\end{displaymath}

The mapping of biological categories of edges onto logical categories
of edges is given in Table \ref{tab:BCEs}.

\FloatBarrier
\newpage
\section{Tables}

\begin{table}[h]
  \caption{Overview of the four custom-built randomization schemes in
    order of their strictness. In particular, the degree of freedeom
    is denoted by the possible pairs of edges available for the
    randomization and by the conserved quantities of the graph:
    source domain (SD), target domain (TD), logical category of an
    edge (LCE), biological category of a vertex (BCV), and biological
    category of an edge (BCE).}\label{tab:randschemes}
  \begin{tabular}{lrl}
    \toprule
    Scheme & Possible pairs & Conserved quantities \\
    \midrule
    DOMAIN      & $134,942,137$ & (SD, TD) \\
    DOMAIN\_LCE &  $59,075,210$ & (SD, TD), LCE \\
    DOMAIN\_BCV &  $54,592,007$ & (SD, TD), BCV \\
    DOMAIN\_BCE &  $42,774,454$ & (SD, TD), BCE \\
    \bottomrule
  \end{tabular}
\end{table}

\begin{table}[h]
  \caption{Summary of the plotted quantities and parameter choices in
    the Figures 2--4 in the main manuscript as well as in the
    Supplementary Figures
    \ref{fig:supp:p099}--\ref{fig:supp:p091}}.\label{tab:params}
  \begin{tabular}{cll}
    \toprule
    Figure & Quantity plotted & Parameter values \\
    \midrule
   2 & $V_t(X, q)$ (or $\vec{\sigma}(t,\,X,q)$) as a function of time
   $t$ & a,c: $X=R$, b,d: $X=M$ \\
   & & a,b: $q = 0.01$, c,d: $q=0.03$ \\
   3 & $\chi^{(R,i)} (X, q) $ as a function of $p = 1-q$ & $i$ in
   $[\text{'unshuffled'},1,2,3,4]$ \textit{(top to bottom)}, \\
   & & $X$ in $[T,R,M]$ \textit{(for each frame)} \\
   4 & $A(X,q)/N$ as function of $p=1-q$ & $X\in\{M,R\}$,
   $i\in\{\text{'unshuffled'},1,2,3,4\}$ \\[0.5em]
   \ref*{fig:supp:p099} & $V_t(X, q)$ (or $\vec{\sigma}(t,\,X,q)$) as
   a function of time $t$ & $X=M,R$; $q=0.01$ \\
   \ref*{fig:supp:p097} &  & $X=M,R$; $q=0.03$\\
   \ref*{fig:supp:p095} &  & $X=M,R$; $q=0.05$\\
   \ref*{fig:supp:p093} &  & $X=M,R$; $q=0.07$\\
   \ref*{fig:supp:p091} &  & $X=M,R$; $q=0.09$\\
   \bottomrule
  \end{tabular}
\end{table}

\begin{table}[htb]
 \caption{Biological categories of a vertex (BVC) of the integrative
   \textit{E.~coli} model and the logical categories of an edge
   (LCEs) a target vertex of this category may contribute to:
   'conjunct' ($C$), 'disjunct' ($D$) and 'regulation' ($R$). The
   detailed vertex composition will be given in
   Grimbs \textit{et al.} \cite{Sgrimbs_integrated_2016}.}\label{tab:BCVs}
 \begin{tabular}{cll}
  \toprule
  Vertex & BCV & LCEs \\
  \midrule
  \tikz\node[gene,fill=myyellow]{}; & gene ($gn$) & $R$ \\
  \tikz\node[protein,fill=myred]{}; & protein monomer ($pm$) & $D$, $R$ \\
  \tikz\node[protein,fill=mygray]{}; & protein-protein-complex
  ($ppc$) & $C$, $D$, $R$ \\
  \tikz\node[protein,fill=myindigo]{}; & protein-compound-complex
  ($pcc$) & $C$, $D$ \\
  \tikz\node[protein,fill=myorange]{}; & protein-rna-complex ($prc$)
  & -- \\
  \tikz\node[reaction,fill=mygreen]{}; & reaction ($rxn$) & $C$, $D$,
  $R$ \\
  \tikz\node[compound,fill=myblue]{}; & compound ($cmp$) & $C$, $D$ \\
  \bottomrule
 \end{tabular}
\end{table}

\begin{table}[htb]
 \caption{Domain-related categories of a vertex (DVCs) of the integrative \textit{E.~coli} model, the corresponding biological categories of vertices (BVCs): \protect\tikz\protect\node[gene,fill=myyellow,scale=0.7]{};~gene, \protect\tikz\protect\node[protein,fill=myred,scale=0.7]{};~protein monomer, \protect\tikz\protect\node[protein,fill=mygray,scale=0.7]{};~protein-protein-complex, \protect\tikz\protect\node[protein,fill=myindigo,scale=0.7]{};~protein-compound-complex, \protect\tikz\protect\node[protein,fill=myorange,scale=0.7]{};~protein-rna-complex, \protect\tikz\protect\node[reaction,fill=mygreen,scale=0.7]{};~reaction, \protect\tikz\protect\node[compound,fill=myblue,scale=0.7]{};~compound, as well as the biological categories of an edge (BCEs) a vertex of this category is involved.}\label{tab:DCVs}
 \begin{tabular}{lll}
  \toprule
  DCV & BCVs & BCEs \\
  \midrule
  gene ($g$) & \tikz\node[gene,fill=myyellow]{}; & $g \rightarrow p$ \\
  protein ($p$) & \tikz\node[protein,fill=myred]{};,
  \tikz\node[protein,fill=mygray]{};,
  \tikz\node[protein,fill=myindigo]{}; & $g \rightarrow p$, $p
  \rightarrow x$ \\
  complex ($x$) & \tikz\node[protein,fill=mygray]{};,
  \tikz\node[protein,fill=myindigo]{}; & $p \rightarrow x$, $c
  \rightarrow x$ \\ enzyme ($z$) & \tikz\node[protein,fill=myred]{};,
  \tikz\node[protein,fill=mygray]{}; & $z \rightarrow r$ \\
  reaction ($r$) & \tikz\node[reaction,fill=mygreen]{}; & $z
  \rightarrow r$, $e \rightarrow r$, $r \rightarrow d$ \\ compound
  ($c$) & \tikz\node[compound,fill=myblue]{}; & $c \rightarrow x$ \\
  educt ($e$) & \tikz\node[protein,fill=myred]{};,
  \tikz\node[protein,fill=mygray]{};,
  \tikz\node[protein,fill=myindigo]{};,
  \tikz\node[compound,fill=myblue]{}; & $e \rightarrow r$ \\
  product ($d$) & \tikz\node[protein,fill=myred]{};,
  \tikz\node[protein,fill=mygray]{};,
  \tikz\node[protein,fill=myindigo]{};,
  \tikz\node[compound,fill=myblue]{}; & $r \rightarrow d$ \\
  \bottomrule
 \end{tabular}
\end{table}

\begin{table}[htb]
 \caption{Biological categories of an edge (BCEs) and the corresponding logical categories of an edge (LCEs): 'conjunct' ($C$), 'disjunct' ($D$) and 'regulation' ($R$), for each vertex linkage of the integrative
   \textit{E.~coli} model. The different linkages are denoted by the combinations of BCVs: \protect\tikz\protect\node[gene,fill=myyellow,scale=0.7]{};~gene, \protect\tikz\protect\node[protein,fill=myred,scale=0.7]{};~protein monomer, \protect\tikz\protect\node[protein,fill=mygray,scale=0.7]{};~protein-protein-complex, \protect\tikz\protect\node[protein,fill=myindigo,scale=0.7]{};~protein-compound-complex, \protect\tikz\protect\node[protein,fill=myorange,scale=0.7]{};~protein-rna-complex, \protect\tikz\protect\node[reaction,fill=mygreen,scale=0.7]{};~reaction, \protect\tikz\protect\node[compound,fill=myblue,scale=0.7]{};~compound. The detailed edge composition will be given in Grimbs \textit{et al.} \cite{Sgrimbs_integrated_2016}.}\label{tab:BCEs}
 \begin{tabular}{lcL{16em}}
  \toprule BCEs & LCEs & Vertex linkages \\
  \midrule
  gene $\rightarrow$ protein ($g \rightarrow p$) & $D$ & 
   \begin{tikzpicture}\node[gene,fill=myyellow](G){};\node[protein,fill=myred](P)at(0.8,0){};\draw[-latex](G)to($(P.west)+(0,0.075)$);\end{tikzpicture} \\
  protein $\rightarrow$ complex ($p \rightarrow x$) & $C$ & 
   \begin{tikzpicture}\node[protein,fill=myred](P){};\node[protein,fill=mygray](C)at(0.8,0){};\draw[-latex]($(P.east)+(0,0.075)$)to($(C.west)+(0,0.075)$);\end{tikzpicture}~ 
   \begin{tikzpicture}\node[protein,fill=myred](P){};\node[protein,fill=myindigo](C)at(0.8,0){};\draw[-latex]($(P.east)+(0,0.075)$)to($(C.west)+(0,0.075)$);\end{tikzpicture}~ 
   \begin{tikzpicture}\node[protein,fill=mygray](P){};\node[protein,fill=mygray](C)at(0.8,0){};\draw[-latex]($(P.east)+(0,0.075)$)to($(C.west)+(0,0.075)$);\end{tikzpicture} \\
  compound $\rightarrow$ complex ($c \rightarrow x$) & $C$ &
   \begin{tikzpicture}\node[compound,fill=myblue](c){};\node[protein,fill=myindigo](P)at(0.8,0){};\draw[-latex](c)to($(C.west)+(0,0.075)$);\end{tikzpicture} \\
  enzyme $\rightarrow$ reaction ($z \rightarrow r$) & $D$ &
   \begin{tikzpicture}\node[protein,fill=myred](P){};\node[reaction,fill=mygreen](R)at(0.8,0){};\draw[-latex]($(P.east)+(0,0.075)$)to(R);\end{tikzpicture}~ 
   \begin{tikzpicture}\node[protein,fill=mygray](P){};\node[reaction,fill=mygreen](R)at(0.8,0){};\draw[-latex]($(P.east)+(0,0.075)$)to(R);\end{tikzpicture} \\
  educt $\rightarrow$ reaction ($e \rightarrow r$) & $C$ &
   \begin{tikzpicture}\node[protein,fill=myred](P){};\node[reaction,fill=mygreen](R)at(0.8,0){};\draw[-latex]($(P.east)+(0,0.075)$)to(R);\end{tikzpicture}~ 
   \begin{tikzpicture}\node[protein,fill=mygray](P){};\node[reaction,fill=mygreen](R)at(0.8,0){};\draw[-latex]($(P.east)+(0,0.075)$)to(R);\end{tikzpicture}~ 
   \begin{tikzpicture}\node[protein,fill=myindigo](P){};\node[reaction,fill=mygreen](R)at(0.8,0){};\draw[-latex]($(P.east)+(0,0.075)$)to(R);\end{tikzpicture}~ 
   \begin{tikzpicture}\node[compound,fill=myblue](c){};\node[reaction,fill=mygreen](R)at(0.8,0){};\draw[-latex](c)to(R);\end{tikzpicture} \\
  reaction $\rightarrow$ product ($r \rightarrow d$) & $D$ &
   \begin{tikzpicture} \node[reaction,fill=mygreen](R){};\node[compound,fill=myblue](c)at(0.8,0){};\draw[-latex](R)to(c);\end{tikzpicture}~ 
   \begin{tikzpicture}\node[reaction,fill=mygreen](R){};\node[protein,fill=myred](P)at(0.8,0){};\draw[-latex](R)to($(P.west)+(0,0.075)$);\end{tikzpicture}~ 
   \begin{tikzpicture}\node[reaction,fill=mygreen](R){};\node[protein,fill=mygray](P)at(0.8,0){};\draw[-latex](R)to($(P.west)+(0,0.075)$);\end{tikzpicture}~ 
   \begin{tikzpicture}\node[reaction,fill=mygreen](R){};\node[protein,fill=myindigo](P)at(0.8,0){};\draw[-latex](R)to($(P.west)+(0,0.075)$);\end{tikzpicture} \\
  transport ($t$) & $C$ &
   \begin{tikzpicture}\node[compound,fill=myblue](c1){};\node[compound,fill=myblue](c2)at(0.8,0){};\draw[-latex,dashed](c1)to(c2);\end{tikzpicture} \\
  regulation ($r^*$) & $R$ &
   \begin{tikzpicture}\node[gene,fill=myyellow](G1){};\node[gene,fill=myyellow](G2)at(0.8,0){};\draw[-latex,dashed](G1)to(G2);\end{tikzpicture}
   
   \begin{tikzpicture}\node[protein,fill=myred](P){};\node[gene,fill=myyellow](G)at(0.8,0){};\draw[-latex,dashed]($(P.east)+(0,0.075)$)to(G);\end{tikzpicture}~ 
   \begin{tikzpicture}\node[protein,fill=myred](P1){};\node[protein,fill=myred](P2)at(0.8,0){};\draw[-latex,dashed]($(P1.east)+(0,0.075)$)to($(P2.west)+(0,0.075)$);\end{tikzpicture}~ 
   \begin{tikzpicture}\node[protein,fill=myred](P1){};\node[protein,fill=mygray](P2)at(0.8,0){};\draw[-latex,dashed]($(P1.east)+(0,0.075)$)to($(P2.west)+(0,0.075)$);\end{tikzpicture}~ 
   \begin{tikzpicture}\node[protein,fill=myred](P){};\node[reaction,fill=mygreen](R)at(0.8,0){};\draw[-latex,dashed]($(P.east)+(0,0.075)$)to(R);\end{tikzpicture} 
  
   \begin{tikzpicture}\node[protein,fill=mygray](C){};\node[gene,fill=myyellow](G)at(0.8,0){};\draw[-latex,dashed]($(C.east)+(0,0.075)$)to(G);\end{tikzpicture}~
   \begin{tikzpicture}\node[protein,fill=mygray](P1){};\node[protein,fill=myred](P2)at(0.8,0){};\draw[-latex,dashed]($(P1.east)+(0,0.075)$)to($(P2.west)+(0,0.075)$);\end{tikzpicture}~ 
   \begin{tikzpicture}\node[protein,fill=mygray](P1){};\node[protein,fill=mygray](P2)at(0.8,0){};\draw[-latex,dashed]($(P1.east)+(0,0.075)$)to($(P2.west)+(0,0.075)$);\end{tikzpicture}~ 
   \begin{tikzpicture}\node[protein,fill=mygray](P){};\node[reaction,fill=mygreen](R)at(0.8,0){};\draw[-latex,dashed]($(P.east)+(0,0.075)$)to(R);\end{tikzpicture}
  
   \begin{tikzpicture}\node[protein,fill=myorange](C){};\node[gene,fill=myyellow](G)at(0.8,0){};\draw[-latex,dashed]($(C.east)+(0,0.075)$)to(G);\end{tikzpicture}~
   \begin{tikzpicture}\node[protein,fill=myorange](C){};\node[protein,fill=myred](P)at(0.8,0){};\draw[-latex,dashed]($(C.east)+(0,0.075)$)to($(P.west)+(0,0.075)$);\end{tikzpicture}
  
   \begin{tikzpicture}\node[protein,fill=myindigo](C){};\node[gene,fill=myyellow](G)at(0.8,0){};\draw[-latex,dashed]($(C.east)+(0,0.075)$)to(G);\end{tikzpicture}
  
   \begin{tikzpicture}\node[compound,fill=myblue](c){};\node[gene,fill=myyellow](G)at(0.8,0){};\draw[-latex,dashed](c)to(G);\end{tikzpicture}~ 
   \begin{tikzpicture}\node[compound,fill=myblue](c){};\node[protein,fill=myred](P)at(0.8,0){};\draw[-latex,dashed](c)to($(P.west)+(0,0.075)$);\end{tikzpicture}~ 
   \begin{tikzpicture}\node[compound,fill=myblue](c){};\node[protein,fill=mygray](C)at(0.8,0){};\draw[-latex,dashed](c)to($(C.west)+(0,0.075)$);\end{tikzpicture}~ 
   \begin{tikzpicture}\node[compound,fill=myblue](c){};\node[reaction,fill=mygreen](R)at(0.8,0){};\draw[-latex,dashed](c)to(R);\end{tikzpicture} \\
  \bottomrule
 \end{tabular}
\end{table}

\FloatBarrier
\clearpage

\section{Sample perturbation trajectories}
Here, further sample trajectories are given similar to the ones in
Fig.~$\ref{fig:states}$ in the main manuscript.

\begin{figure}[h]
  \centering
  \includegraphics[width=.9\columnwidth]
                  {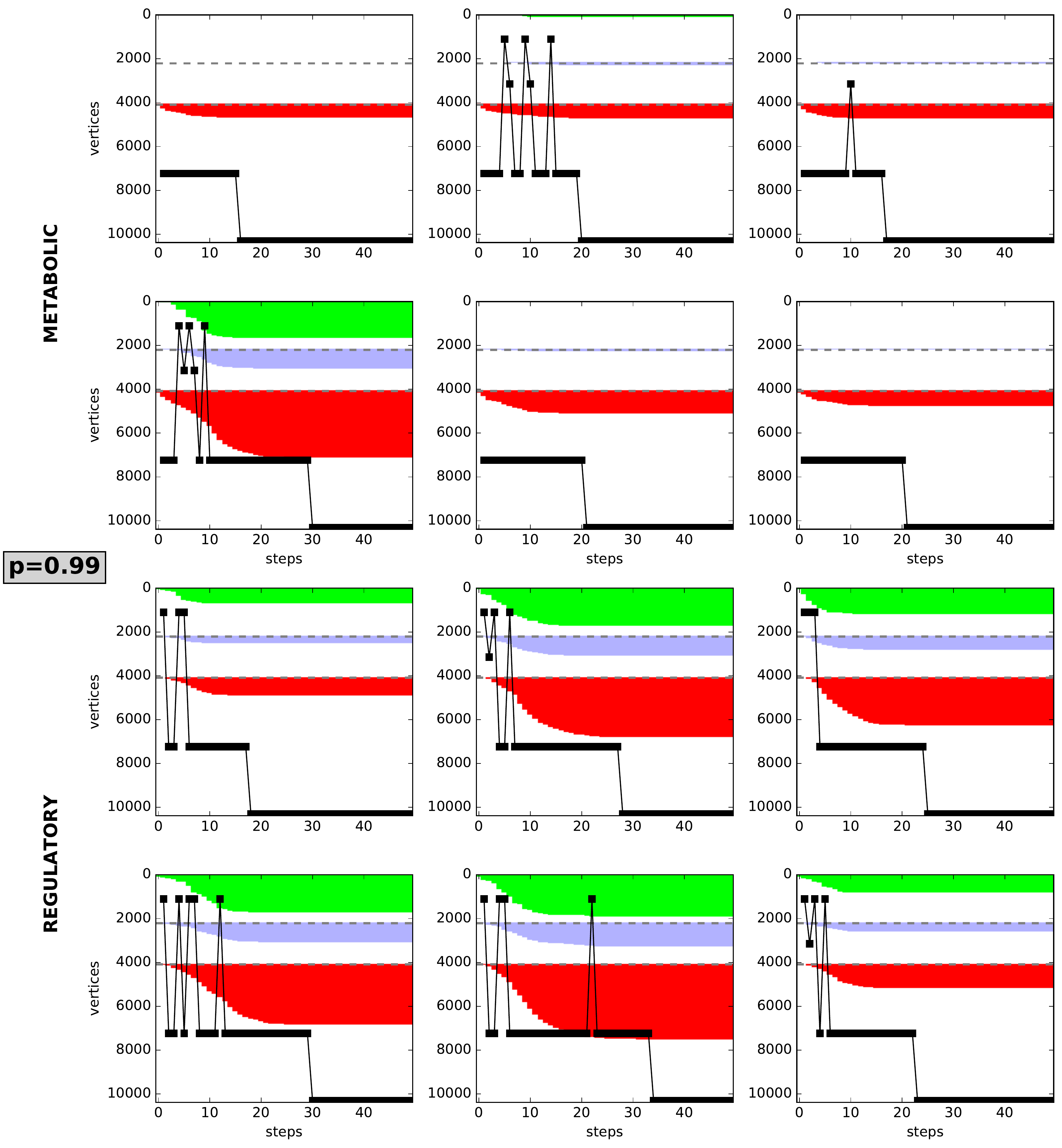}
  \caption{\label{fig:supp:p099} Six sample trajectories for single
    perturbations of size $q=0.01$ in the metabolic \textit{top two
      rows} and regulatory domain \textit{two bottom rows},
    respectively.}
\end{figure}
\begin{figure}[h]
  \centering
  \includegraphics[width=.9\columnwidth]
                  {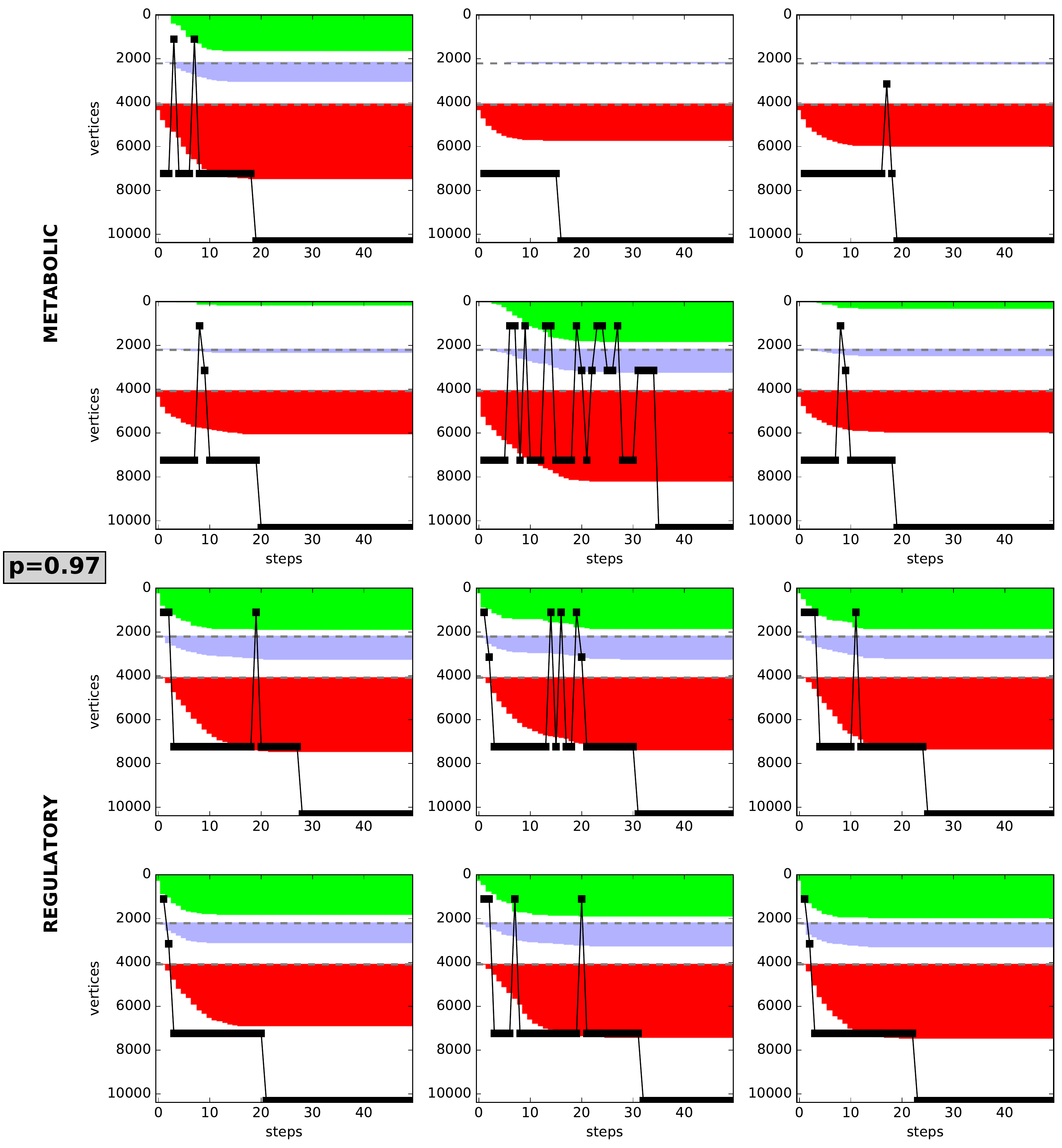}
  \caption{\label{fig:supp:p097} Six sample trajectories for single
    perturbations of size $q=0.03$ in the metabolic \textit{top two
      rows} and regulatory domain \textit{two bottom rows},
    respectively.}
\end{figure}
\begin{figure}[h]
  \centering
  \includegraphics[width=.9\columnwidth]
                  {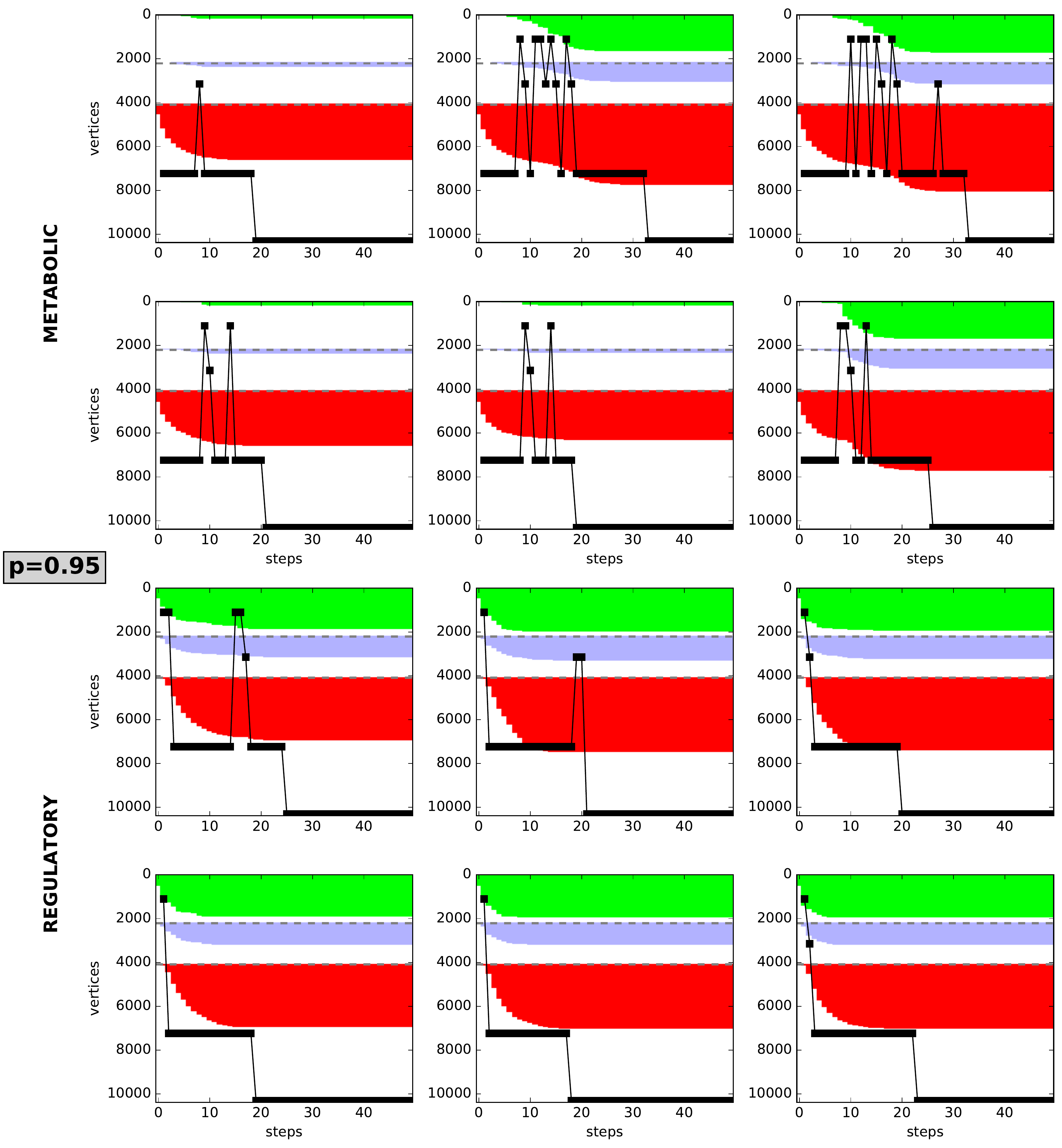}
  \caption{\label{fig:supp:p095} Six sample trajectories for single
    perturbations of size $q=0.05$ in the metabolic \textit{top two
      rows} and regulatory domain \textit{two bottom rows},
    respectively.}
\end{figure}
\begin{figure}[h]
  \centering
  \includegraphics[width=.9\columnwidth]
                  {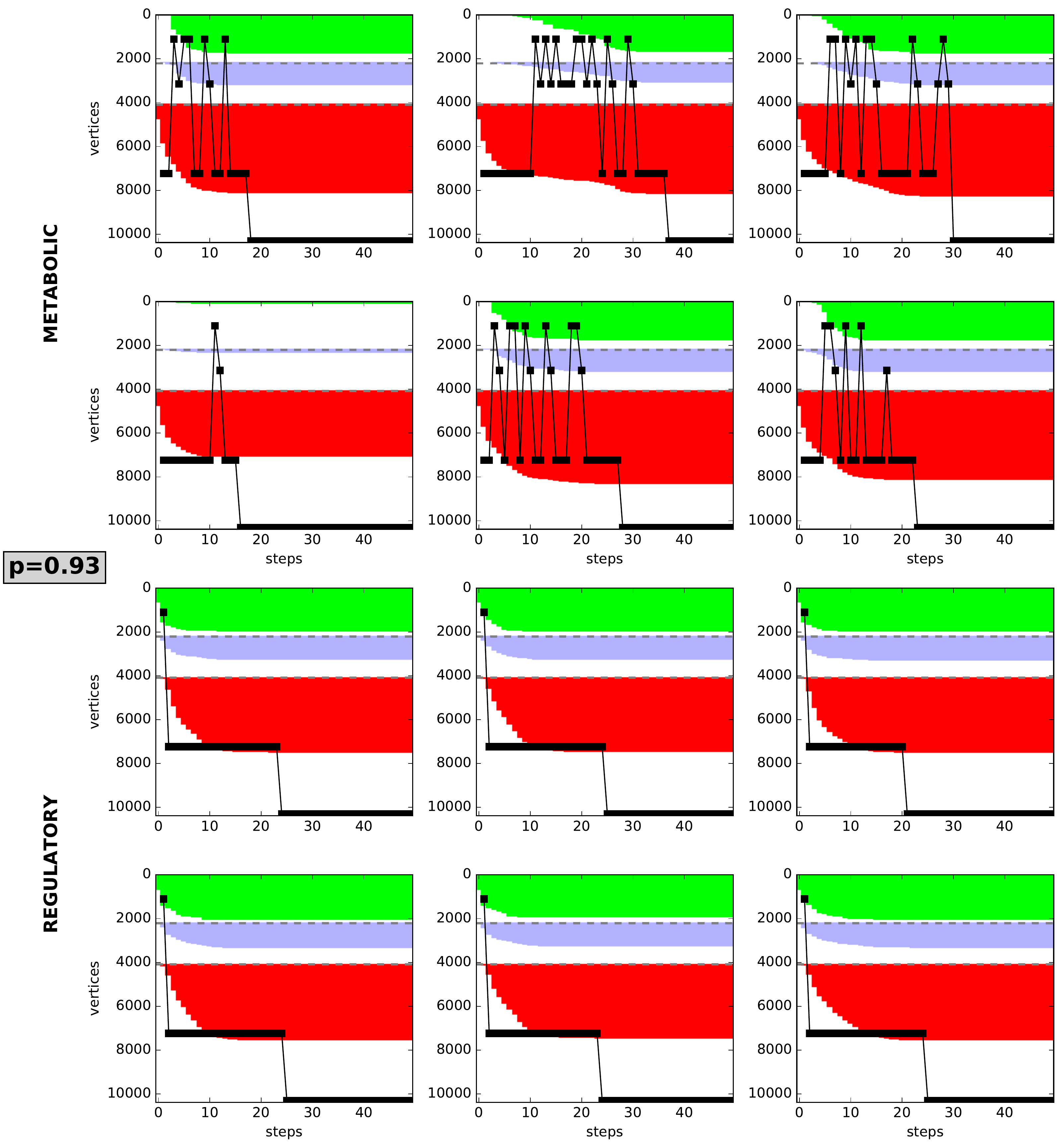}
  \caption{\label{fig:supp:p093}Six sample trajectories for single
    perturbations of size $q=0.07$ in the metabolic \textit{top two
      rows} and regulatory domain \textit{two bottom rows},
    respectively.}
\end{figure}
\begin{figure}[h]
  \centering
  \includegraphics[width=.9\columnwidth]
                  {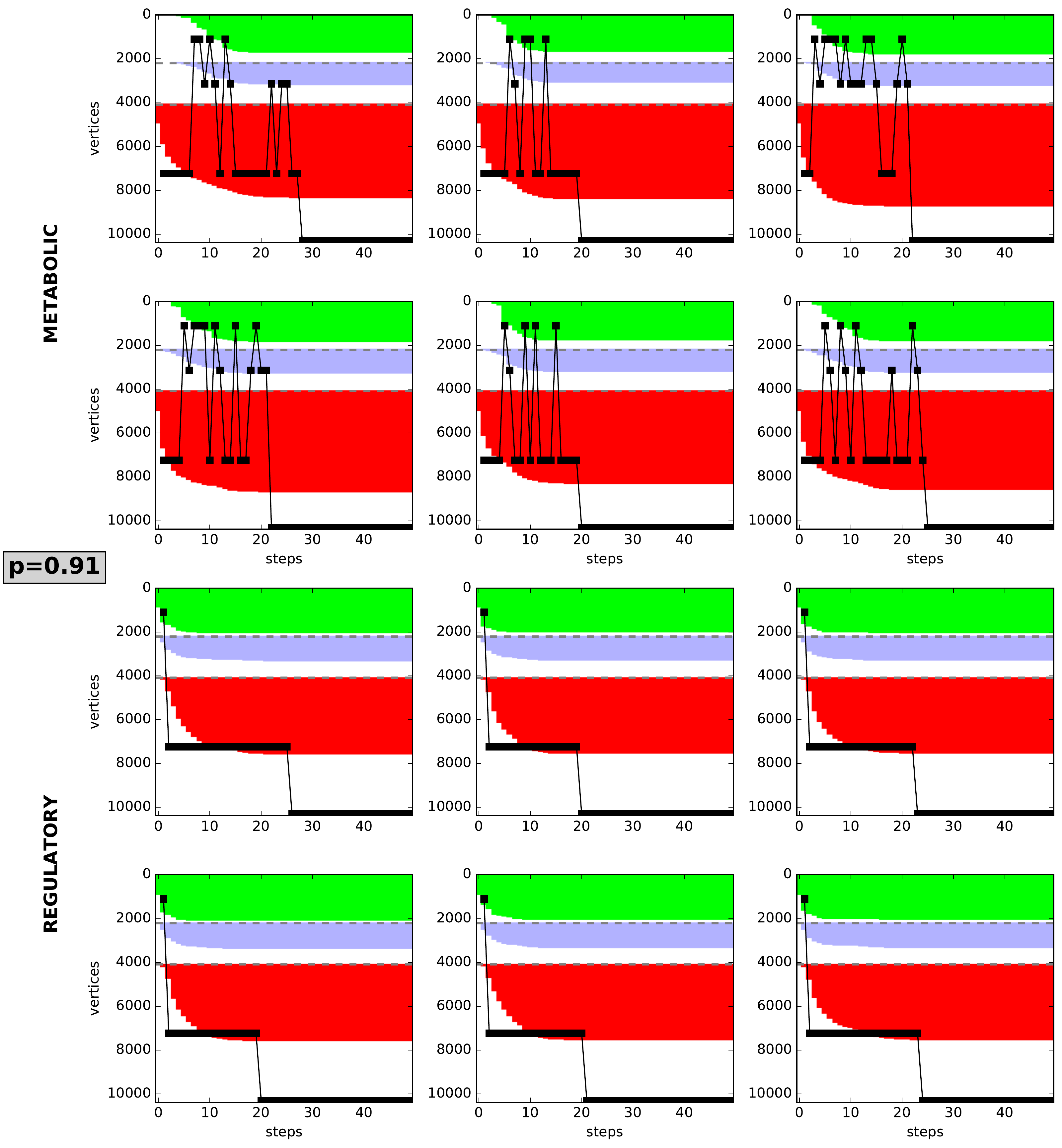}
  \caption{\label{fig:supp:p091}Six sample trajectories for single
    perturbations of size $q=0.09$ in the metabolic \textit{top two
      rows} and regulatory domain \textit{two bottom rows},
    respectively.}
\end{figure}

\FloatBarrier
\newpage

\end{document}